\documentclass[final,3p,times,10pt]{elsarticle}

\usepackage{amssymb}
\usepackage{amsthm}
\usepackage{amsmath}
\usepackage{amsfonts}
\usepackage{graphicx}
\usepackage{mathrsfs}
\usepackage{latexsym}
\usepackage{graphics}
\usepackage{epsf}
\usepackage{epsfig}
\usepackage{subfigure}
\usepackage{bm}
\usepackage{subfigmat}
 \usepackage{url} 
\usepackage{psfrag}

\begin{document}

\begin{frontmatter}



\title{Detached eddy simulation of shock unsteadiness in an over-expanded planar nozzle  }

\author[a]{E. Martelli\corref{cor1}}
\ead{emanuele.martelli@unina2.it}
   \author[b]{P.P. Ciottoli} \author[b]{M. Bernardini} \author[b]{F. Nasuti}  \author[b]{M. Valorani} 
\address[a]{Second University of Naples, 81031 Aversa, Italy}
\address[b]{Univerity of Rome "La Sapienza", 00184 Rome, Italy}
\cortext[cor1]{Corresponding author}

\begin{abstract}
This work investigates the self-excited shock wave oscillations in a three-dimensional planar over-expanded nozzle turbulent flow by means of Detached Eddy Simulations. Time resolved wall pressure measurements are used as primary diagnostics. The statistical 
analysis reveals that the  shock unsteadiness has common features in terms of the root mean square of the pressure fluctuations with other
classical shock wave/boundary layer interactions, like compression ramps 
and incident shocks on a flat plate.  
The Fourier transform and the continuous wavelet transform are used to conduct the spectral analysis.
The results of the former indicate that the pressure in the shock region is characterized by a broad low-frequency content, without any resonant tone.  
The wavelet analysis, which is well suited to study non stationary process, reveals that the pressure signal is characterized by an amplitude and a frequency modulation in time.
\end{abstract}

%

\end{frontmatter}


\section{Introduction}
\label{}

During the sea-level start-up of a liquid rocket engine the nozzle is highly overexpanded and an internal flow separation takes place,
characterized by a shock wave boundary layer interaction (SWBLI), which causes the
shedding of vortical structures and unsteadiness in the shock wave position.
In the nozzle design community, flow separation is considered dangerous, since it produces dynamic side-loads that reduce the safe life of the engine and can lead to a failure of the nozzle structure.
The need to improve nozzle performance under overexpanded conditions and to mitigate the side loads fostered several experimental~\cite{nave,ostlund02,nguyen} and numerical investigations~\cite{chen,nasuti98,deck02}.  All these studies revealed two distinct
separation processes: the free shock separation (FSS), in which the boundary layer separates from the nozzle wall and never reattaches, and the restricted shock separation (RSS), characterized by a closed recirculation bubble and the reattachment of the shear layer to the wall.

According to Schmucker~\cite{Schmucker}, the main cause of side-loads appearance is an asymmetry of the separation location, which produces a tilted separation line, a momentum imbalance and consequently a lateral force.
Nave and Coffey~\cite{nave} observed that, in optimized nozzles, the maximum value of the side loads takes place during the transition from  FSS to  RSS condition.
A literature survey reporting the various studies on the side loads generation and separation shock configurations can be found in Hadjadj and Onofri~\cite{hadjadj09} and in Reijasse et al.~\cite{Reijasse}.
But, in spite of all of these studies, a fundamental knowledge of supersonic flow physics in the presence of a shock separation interaction is still needed.

One of the several tasks for future investigations recommended by Hadjadj and Onofri~\cite{hadjadj09} is related to the low frequency oscillations of a shock interacting with a turbulent flow separation.
This phenomenon, consisting in fluctuating pressure loads and pulsating separated flows, should be carefully considered by rocket nozzle designers.
A lot of experimental work has been carried out to understand the unsteadiness of shocks in internal flows.
Bogar et al.~\cite{Corp1981} investigated the unsteady flow characteristics of a supercritical transonic diffuser as a function of shock Mach number and diffuser length.
They observed that in the case of attached flow (or very mild separation), the characteristic  frequencies have an acoustic nature,
and scale with the distance of the shock from the diffuser exit. While, in the case of separated flows, the characteristic frequencies  scale with the length of the inviscid core flow.
Zaman et al\cite{Zaman2002} carried out experiments in a supersonic planar diffuser, and documented a transonic tone with weak harmonics. The presence of harmonics led to the theory that the observed frequency was caused by an acoustic feed-back mechanism.

In addition, they found that the instability has the higher sound pressure level if the boundary layer before the shock is laminar, while if the boundary layer is tripped, in some of the tests the tone was suppressed. Both Bogar and Zaman underlined that the mechanism for the shock instability is unclear, although they indicate some dependency of the shock dynamics on the downstream separated region.
Also Handa et al. \cite{Handa2003}, in their experimental investigation of a transonic diffuser, underlined two different mechanisms for the shock oscillation. In the first mechanism, pressure disturbances, generated in the downstream turbulent separated region, force the shock to oscillate, resulting in a broad shape of the power spectral density. In addition, the intensity of this movement is mainly governed by the Mach number in front of the shock.
The second mechanism foresees the reflection of a disturbance at the diffuser exit (acoustic feedback), resulting in a narrow-shaped power spectral density.
More recently, Johnson and Papamoschou~\cite{Johnson2010} have studied the unsteady shock behavior in an  over-expanded planar nozzle.
Their results indicate a low frequency piston-like shock motion without any resonant tones.
Correlations of Pitot pressure with wall pressure  indicated a strong coherence of the shear layer instability with the shock motion.
All these different experimental investigations suggest a correlation between the shock movement and the acoustic or fluid dynamic characteristics downstream of the flow separation.

As far as Large/Detached Eddy Simulations (LES/DES) of this kind of flows are concerned, very few studies can be found in literature. Deck~\cite{Deck2009} carried out a Delayed Detached Eddy Simulation (DDES) of the end-effect regime in an axi-symmetric over-expanded rocket nozzle flow. While the experimentally measured main properties of the flow motion were rather well reproduced, the computed main frequency resulted to be higher than in the experiment.
Olson and Lele~\cite{Olson2013}  performed large eddy simulations of the experiments of Papamoschou, finding a satisfactory agreement between  the experimental data and the computed frequency  of the shock displacement.
The origin of the unsteadiness was attributed to the confinement of exit area by the separated flow.

In the present work, Delayed Detached Eddy Simulations (DDES) reproducing the flowfield of the nozzle geometry of Bogar et al.~\cite{Corp1981} are carried out to completely characterize the shock unsteadiness at different nozzle pressure ratios.
Resorting to this computational strategy, any perturbation coming from the upstream boundary layer is not resolved. Therefore the separation shock can
only be perturbed by the turbulent recirculating region. In a recent review of Clemens et al. \cite{Clemens2014} on the low frequency unsteadiness of the shock wave/turbulent boundary layer interaction,
it is argued that both upstream and downstream perturbations are at work in the interaction, whether there is a separation bubble or not.
However,  the influence of the upstream turbulent boundary layer decreases when the size of the separation bubble   increases. In flow separated nozzle, the recirculation bubble is longer than in other classical configuration like the compression ramp. Therefore it seems reasonable to attribute the most important influence on the shock movement to the downstream region. In such a case, the DDES technique can be considered adequate.
Following the literature~\cite{Bamlaart}, the unsteady behavior of shock flow separation interaction is characterized by analyzing the time series of the wall pressure signals. In particular, the stream-wise distributions
of the root mean square as well as the probability density functions (pdf) of the pressure signals are discussed.
Then the intermittency of the signals is computed, and from its distribution the length scale of the shock motion is quantified, this quantity being directly linked to the level of the side-loads.

The shock flow separation interaction is characterized by the presence of several dominating components in the  pressure  fluctuations, connected to the development of vortical structures and to the shock movement itself.
Fourier analysis is used in order to evaluate the main frequencies along the nozzle, from the shock location to the turbulent recirculating zone.
In addition to the global spectral characterization,  it is of great interest to understand
if the dominating components are simultaneously or alternatively present, and to characterize the time variation of their frequency and amplitude.
The wavelet transform is an analysis tool well suited to the study of multi-scale, non-stationary processes occurring over finite temporal domains.
In particular, it allows to detect the localized variations of power within a time series.
In fact, by decomposing a time series into time-frequency space, it is possible to determine both the dominant modes of variability,  as well as their time evolution.  Therefore, the continuous complex wavelet transform is directly applied in order to identify the time modulation of frequency and amplitude of turbulent structures \cite{Farge}.
This characterization contributes to the basic understanding of the shock wave/turbulent boundary layer interaction physics in internal flows. A better knowledge of this phenomenon can help in predicting and possibly control the level of side loads, thus allowing the design  of larger area ratio nozzle, improving the nozzle performance and reducing the costs of the access to space.

The paper is organized as follows. In section~\ref{sec:Compsetup} the numerical method is
introduced and the use of DDES is discussed in the frame of over-expanded nozzles.
Section~\ref{sec:TestCase} presents the experimental test cases and the numerical setup.
Section~\ref{sec:ResDisc} presents the mean properties of the fields and of the wall pressure distributions,
together with the statistical description of the shock separation iteration.
Fourier and the wavelet spectral analysis are discussed in Section~\ref{sec:WallPSig}
and Section~\ref{sec:SpecAna} respectively. Finally, in the conclusion section the major
findings of this investigation are reported.

\section{Computational setup}
\label{sec:Compsetup}

To better understand the unsteadiness of SWBLI in supersonic nozzles and the role played in the generation of side loads,
large eddy simulations should be ideally carried out to capture the larger structures of the turbulent flow.
Unfortunately, the computational cost of a pure (wall-resolved) LES is still very high for high-Reynolds number wall-bounded flows.
To overcome this limitation, hybrid RANS/LES modeling approaches have been proposed to simulate massively separated flows,
such as the well-known DES \cite{Spalart2009}.
A general feature of this approach is that the whole or at least a major part of the attached boundary layer is
treated resorting to RANS, while LES is applied only in the separated flow regions.
In the following a brief description of the numerical solver used is given, then the main characteristics of
DDES are outlined.

\subsection{Physical model} 

We solve the three-dimensional Navier-Stokes equations for a compressible, viscous, heat-conducting gas
\begin{equation}
 \left .
 \begin{aligned}
   \frac{\partial \rho}{\partial t} + \frac{\partial (\rho \, u_j)}{\partial x_j}
     & = 0, \\
   \frac{\partial (\rho \, u_i)}{\partial t} + \frac{\partial (\rho \, u_i u_j)}{\partial x_j} +
       \frac{\partial p}{\partial x_i} - \frac{\partial \tau_{ij}}{\partial x_j} & = 0, \\
   \frac{\partial (\rho \, E)}{\partial t} + \frac{\partial (\rho \, E u_j + p u_j)}{\partial x_j}
     - \frac{\partial (\tau_{ij} u_i - q_j)}{\partial x_j} & = 0, \label{eq:ns}
 \end{aligned}
 \right .
\end{equation}
where $\rho$ is the density, $u_i$ is the velocity component in the $i$-th coordinate direction ($i=1,2,3$),
$E$ is the total energy per unit mass, $p$ is the thermodynamic pressure.
The total stress tensor $\tau_{ij}$ is the sum of the viscous and the Reynolds stress tensor, 
\begin{equation}
 \label{eq:stress}
 \tau_{ij} = 2 \, \rho \left ( \nu + \nu_t \right ) S^*_{ij} \qquad S^*_{ij} = S_{ij} - \frac 13 \, S_{kk} \, \delta_{ij},
\end{equation}
where the Boussinesq hypothesis is applied through the introduction of the eddy viscosity $\nu_t$, and 
%
%
where $S_{ij} = \left( u_{i,j} + u_{j,i} \right) /2 $ is the strain-rate tensor, $\nu$
the molecular viscosity, depending on temperature $T$ through Sutherland's law.
Similarly, the total heat flux $q_j$ is the sum of a molecular and a turbulent contribution
\begin{equation}
 q_j = -\rho \, c_p \left ( \frac{\nu}{\mathrm{Pr}} + \frac{\nu_t}{\mathrm{Pr}_t} \right ) \frac{\partial T}{\partial x_j},
\end{equation}
$\mathrm{Pr}$, $\mathrm{Pr}_t$ being the molecular and turbulent Prandtl numbers, assumed to be
0.72 and 0.9, respectively.
Hybrid RANS/LES capabilities are provided through the implementation of the delayed detached-eddy simulation (DDES) approach
based on the Spalart-Allmaras (SA) model~\cite{spalart06}, which involves a transport equation for a pseudo eddy viscosity $\tilde{\nu}$ 
\begin{equation} 
 \label{eq:sa}
 \frac{\partial (\rho \tilde{\nu})}{\partial t} + \frac{\partial (\rho \, \tilde{\nu} \,u_j)}{\partial x_j} = 
 c_{b1} \tilde{S} \rho \tilde{\nu} +
 \frac{1}{\sigma}
 \left [
  \frac{\partial}{\partial x_j} \left [ \left ( \rho \nu + \rho \tilde {\nu} \right ) \frac{\partial \tilde{\nu}}{\partial x_j} \right ] +
  c_{b2} \, \rho \left ( \frac{\partial \tilde{\nu}}{\partial x_j} \right )^2
 \right ]
  -c_{w1}  f_w \rho \left ( \frac{\tilde{\nu}}{\tilde{d}} \right )^2,
\end{equation}
where $\tilde{d}$ is the model length scale, $f_w$ is a near-wall damping function, $\tilde{S}$
a modified vorticity magnitude, and $\sigma, c_{b1}, c_{b2}, c_{w1}$ model constants.
The eddy viscosity in Eq.~\ref{eq:stress} is related
to $\tilde{\nu}$ through $\nu_t = \tilde{\nu} \, f_{v1}$, where $f_{v1}$ is a
correction function designed to guarantee the correct boundary-layer behavior in the near-wall region.
In DDES the destruction term in Eq.~\ref{eq:sa} is designed so that the model reduces to pure RANS in attached boundary layers and
to a LES sub-grid scale one in the detached flow regions.
This is accomplished by defining the length scale $\tilde{d}$ as
\begin{equation}
 \tilde{d} = d_w - f_d \, \textrm{max} \left (0, d_w-C_{DES} \, \Delta \right),
\end{equation}
where $d_w$ is the distance from the closest wall, $\Delta$ is the subgrid length-scale, controlling the wavelengths resolved in LES mode.
The function $f_d$, designed to be $0$ in boundary layers and $1$ in LES regions, reads as
\begin{equation}
 f_d = 1-\tanh{\left [ \left ( 8 r_d \right )^3 \right ]}, \qquad r_d = \frac{\tilde{\nu}}{k^2 \, d_w^2 \, \sqrt{U_{i,j} U_{i,j}}},
\end{equation}
where $U_{i,j}$ is the velocity gradient and $k$ the von Karman constant.
The introduction of $f_d$ distinguishes DDES from the original DES approach~\cite{spalart97} (usually denoted
as DES97), ensuring that boundary layers are treated in RANS mode also in the presence of ``ambiguous'' grids in the sense defined
by Spalart et al.~\cite{spalart06}, for which the wall-parallel spacings do not exceed the boundary layer thickness.
The DDES strategy prevents the phenomenon of model stress depletion, consisting in the excessive reduction
of the eddy viscosity in the region of switch (grey area) between RANS and LES, which in turn leads to
grid-induced separation. 
Unlike in the original DDES formulation, the sub-grid length scale in this work is not defined
as the largest spacing in all coordinate directions $\Delta_{\mathrm{max}} = \max (\Delta x, \Delta y, \Delta z)$,
but it depends on the flow itself, through $f_d$ as follows
\begin{equation}
 \label{eq:delta}
 \Delta = \frac{1}{2} 
 \left [
 \left (1+\frac{f_d-f_{d0}}{|f_d-f_{d0}|} \right ) \, \Delta_{\textrm{max}} +
 \left (1-\frac{f_d-f_{d0}}{|f_d-f_{d0}|} \right ) \, \Delta_{\textrm{vol}}
 \right ],
\end{equation}
where $f_{d0} = 0.8$, $\Delta_{\mathrm{vol}} = (\Delta x \cdot \Delta y \cdot \Delta z)^{1/3}$.
The improvement over the classical $\Delta_{max}$ definition is shown in Deck~\cite{deck12}, where
the problem of the delay in the formation of flow instabilities encountered in early applications
of DES/DDES is solved.

\subsection{Numerical method} 

Numerical simulations are carried out by means of a in-house, fully validated
compressible flow solver, that exploits a centered second-order finite volume
approach and takes advantage of an energy consistent formulation (away from shocks). Cell-face values
of the flow variables are obtained from the cell-centered values through suitable reconstructions. In smooth flow regions, the reconstruction
is carried out in such a way that the overall kinetic energy of the fluid is preserved, in the limit of inviscid, incompressible flow~\cite{Pirozzoli2011}.
This property is particularly beneficial for
flow regions treated in LES mode, where the grid is sufficiently fine
to support the development of LES content, and where the only relevant dissipation (in addition
to the molecular one) should be that provided by the turbulence model.
The discretization scheme is made to switch to third-order weighted essentially-non-oscillatory (WENO) near discontinuities,
as controlled by a modified Ducros sensor~\cite{ducrosetal99}. The gradients normal to the cell faces needed for the viscous fluxes,
are evaluated through second-order central-difference approximations,
obtaining compact stencils and avoiding numerical odd-even decoupling phenomena.
Time advancement of the semi-discretized system of ODEs resulting from the spatial discretization is carried out by
means of a low-storage third-order Runge-Kutta algorithm~\cite{Bernardini2009}.
The code is written in Fortran 90, it uses domain decomposition and it fully exploits the Message Passing Interface (MPI) paradigm for the parallelism.

\section{Test case description}
\label{sec:TestCase}
\begin{figure}
\centering
 {\includegraphics[width=0.70\textwidth]{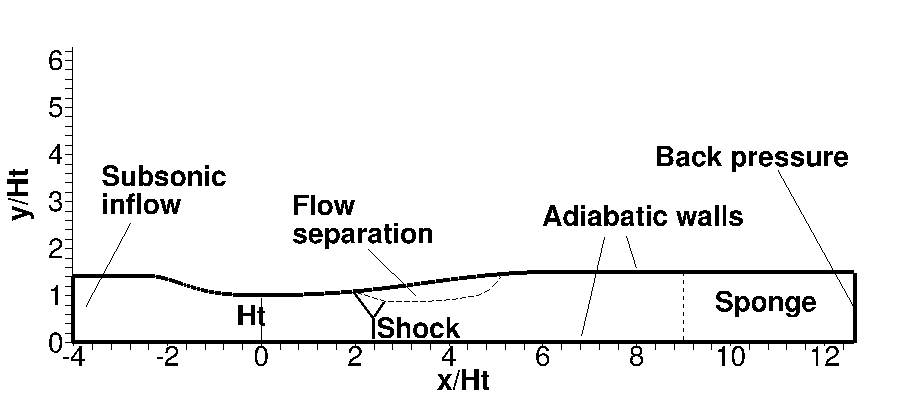}}
 \caption{Two-dimensional schematic of the computational domain with boundary conditions} 
 \label{f:comp}
\end{figure}

The experimental diffuser model numerically reproduced in this work is a convergent-divergent channel with a flat bottom and a contoured top wall, as shown in figure~\ref{f:comp}. 
The analytical expression of the contoured wall can be found in Bogar et al.~\cite{Corp1981} 
The channel height at the throat is 44 mm, the exit-to-throat area ratio is 1.52, the throat cross-sectional aspect ratio is 4.0,
and the divergent length to throat height ratio is 7.2. For $x/H_t$ greater than 7.2 the nozzle is characterized by a constant area section.
In the experimental case dry air is supplied to the model from a plenum chamber immediately upstream. 
The flow from the model is vented to the atmosphere, providing a constant-pressure downstream boundary condition.

A two-dimensional schematic of the computational domain adopted in the simulations is presented in figure~\ref{f:comp}. According to Bogar et al.~\cite{Corp1981} an arbitrarily selected location within the constant area section upstream of the throat ($x_i/H_t=-4.04$) has been chosen as a nominal inlet section. 
The nominal exit section is at $x_e/H_t=12.63$. In the inlet station a subsonic flow is prescribed by imposing the Mach number $M=0.46$, the static pressure and 
the flow direction. The top and bottom surfaces are treated as adiabatic no-slip walls. In the spanwise direction the extent of the domain is $L_z/H_t=4$, and periodic boundary conditions are imposed. 
At the exit section a characteristics based boundary condition prescribing the back pressure is assigned. In order to avoid any acoustic coupling a sponge is imposed from the station at $x/H_t=9$. As stated in the introduction, in fact, the effect of the acoustic feed back is less important when a large recirculating bubble is present.
A logically Cartesian structured mesh is generated using the conformal mapping algorithm of Driscoll and Vavasis\cite{Driscoll1998} and the open-source tool gridgen-c. The computational mesh consists in $N_x\cdot N_y\cdot N_z= 512\cdot192\cdot256$ cells for a total number of $N_{xyz} \approx 25\cdot 10^6 $ cells.   
 
\begin{figure}
\centering
 {\includegraphics[width=0.50\textwidth]{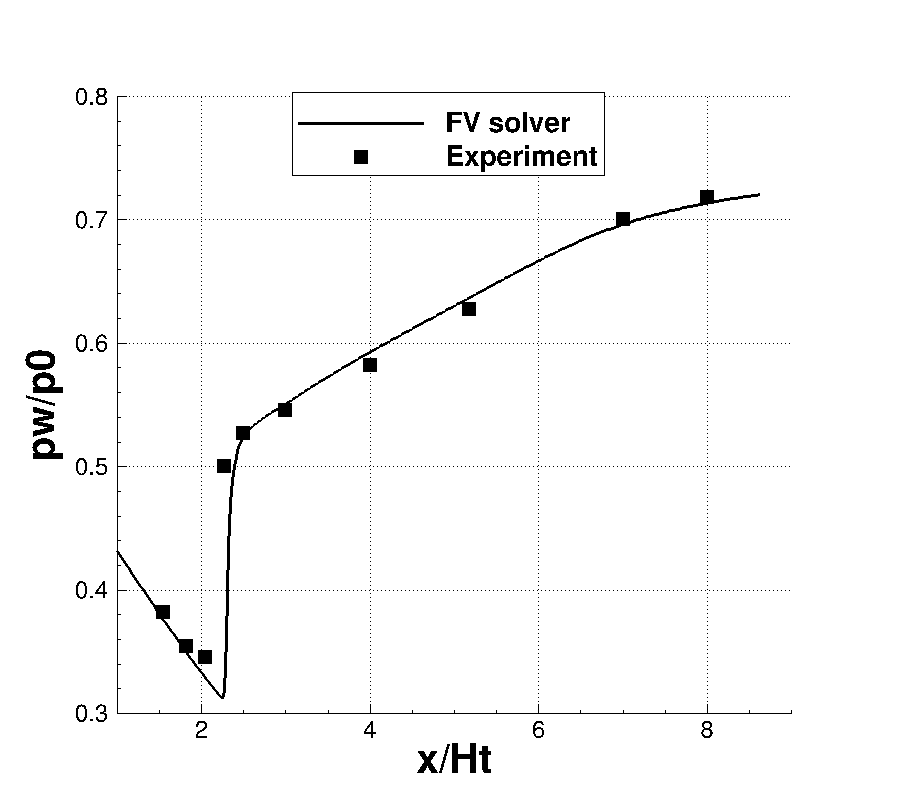}}
 \caption{Experimental mean top wall pressure distribution~\cite{Corp1981}, compared with a 2D RANS simulation for NPR = 1.39} 
 \label{f:cfr-pw-exp}
\end{figure}


The present internal flowfield is characterized by the nozzle pressure ratio $NPR = p_0/p_a$,  
where $p_0$ and $p_a$ denote the chamber and ambient pressure respectively. In this work three different values are simulated: 1.39, 1.46 and 1.54.  
The NPR's values are changed by decreasing the back pressure at the exit section, so that the nozzle Reynolds number based on the chamber values and the throat height remains constant:
$$Re = \frac{\sqrt{\gamma}}{\mu}\frac{p_cH_t}{\sqrt{R_{air}T_0}}=1.5 \cdot 10^6,$$ 
where $\gamma$ is the constant specific heat ratio, $\mu$ is the molecular viscosity evaluated at the chamber temperature $T_0$  and $R_{air}$ is the air gas constant.
A preliminary 2D RANS simulation was performed in order to verify that the mesh resolution in the longitudinal and wall normal direction is sufficient to capture the position of separation line. The comparison of the computed top wall pressure distribution with the experiment~\cite{Corp1981} for NPR = 1.39 is reported in figure~\ref{f:cfr-pw-exp}, and it shows that 
both the position of the separation point and the pressure  behavior in the separated zone are well reproduced.

\section{Results and discussion}

\subsection{Time-averaged and instantaneous flowfield} 
\label{sec:ResDisc}
\begin{figure}
 \begin{subfigmatrix}{2}
  \subfigure[]{\includegraphics[width=0.40\textwidth]{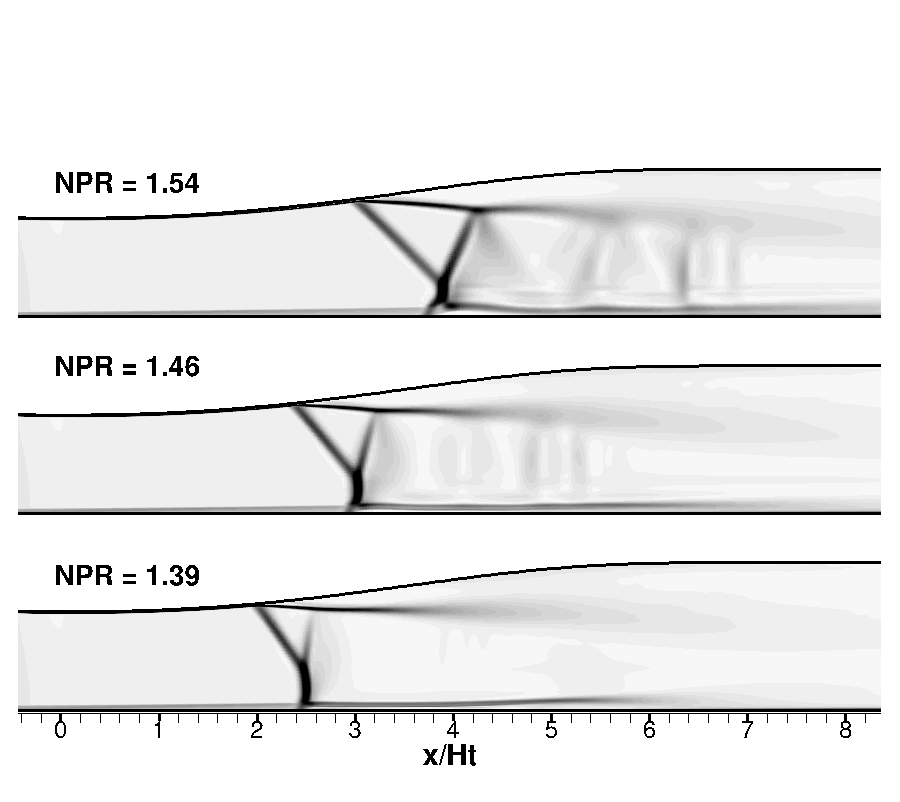}}
  \subfigure[]{\includegraphics[width=0.40\textwidth]{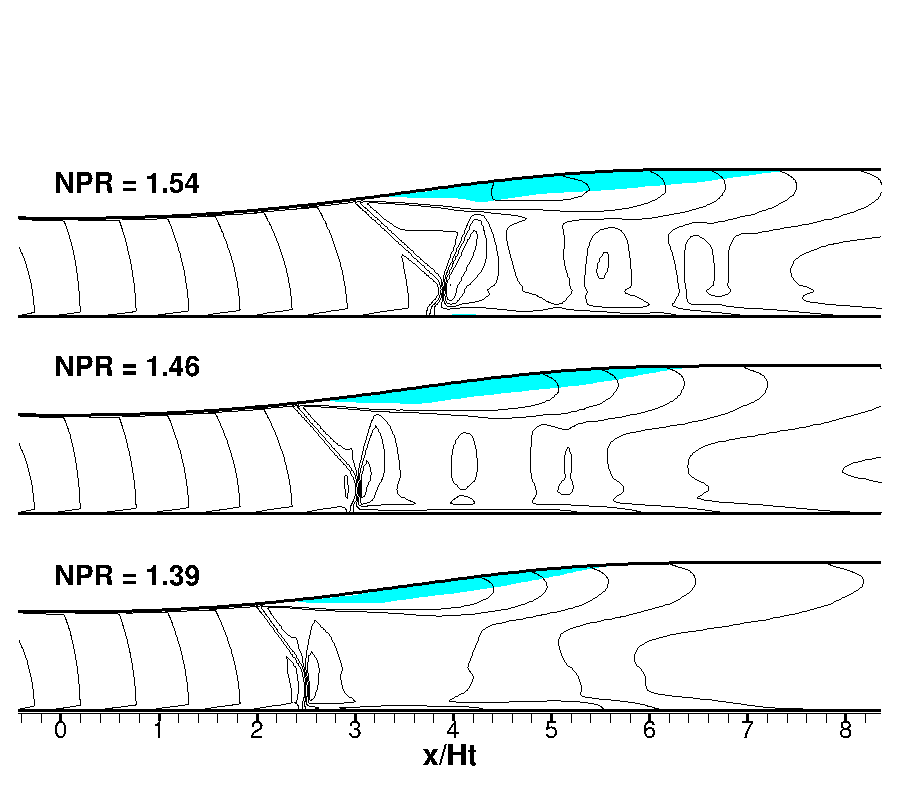}}
 \end{subfigmatrix}
 \caption{Time-averaged and span-wise averaged  field of: a)  $||\nabla \rho||$ and b) density (isolines) with the region with negative velocities (coloured) in the transonic nozzle at different NPR's.}
 \label{f:grad_rho}
\end{figure}

The time and spanwise averaged numerical Schlieren like visualizations ($||\nabla \rho||$) of the three NPR's are presented in Figure~\ref{f:grad_rho}. The flowfield is characterized by a
lambda shock, a recirculation zone and a shear layer. As the NPR increases, the separation shock moves downstream, while the height of the Mach stem decreases. 
The evolution of the recirculation bubble with the NPR is shown in figure~\ref{f:grad_rho}b where an enlargement of the back flow region (indicated by the coloured flood region) with the downstream movement of the separation shock can be observed. 
\begin{figure}
 \begin{subfigmatrix}{2}
  \subfigure[]{\includegraphics[width=0.40\textwidth]{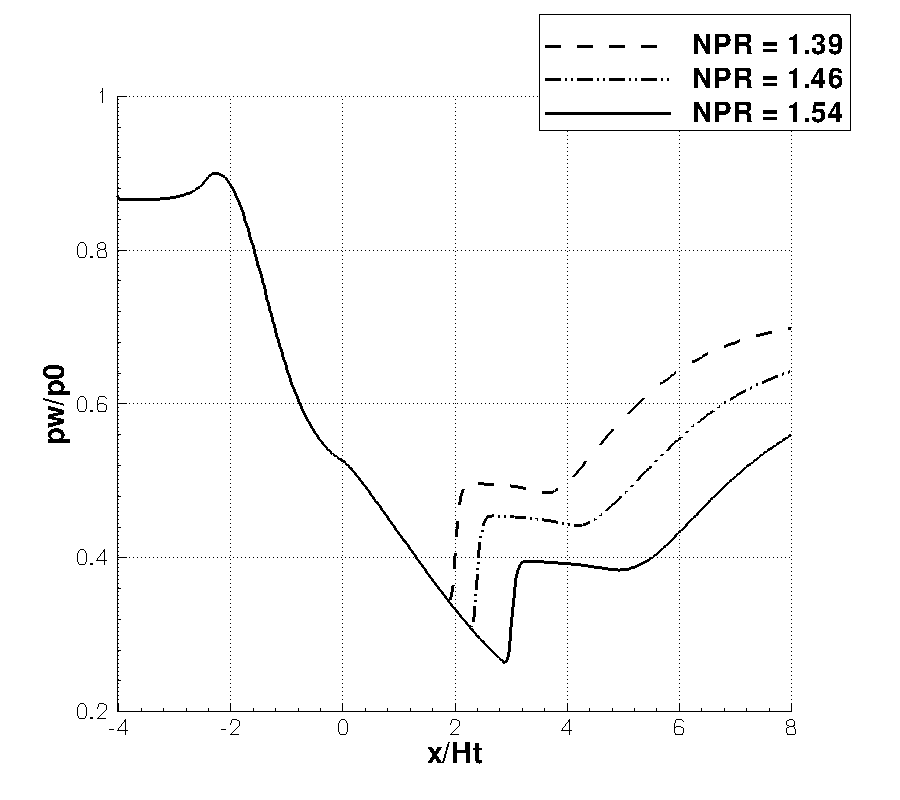}}
  \subfigure[]{\includegraphics[width=0.40\textwidth]{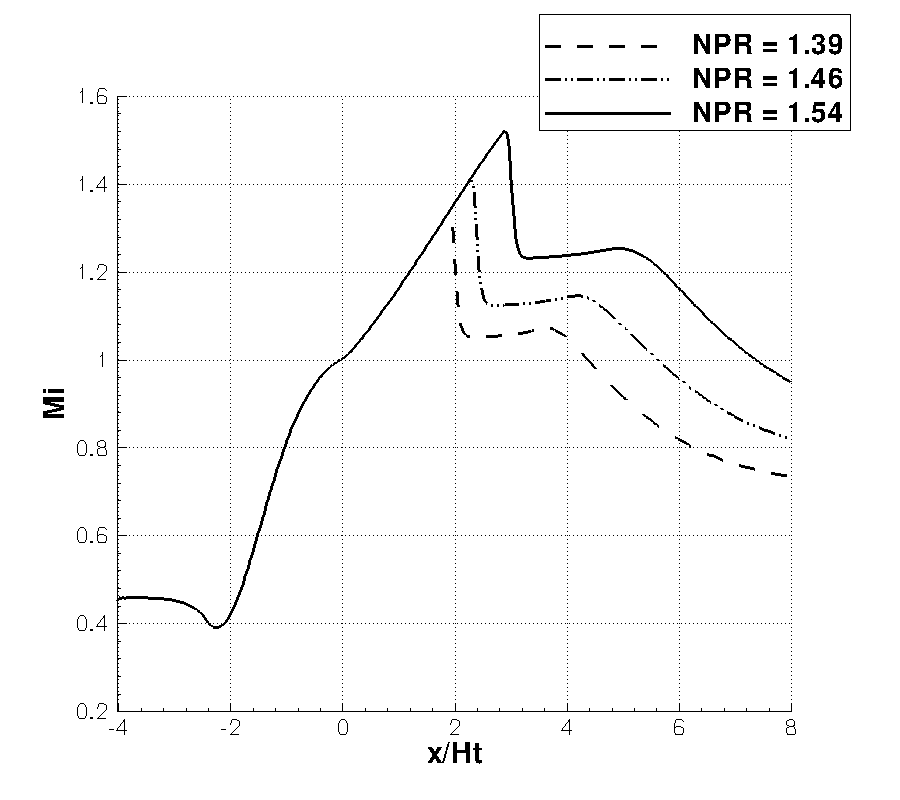}}
 \end{subfigmatrix}
 \caption{Left: streamwise distributions of time and spanwise averaged wall pressure. Right: streamwise distributions of the wall isentropic Mach number, indicating the shock strength}
 \label{f:pw-mach}
\end{figure}

The top wall pressure distributions in the streamwise direction are presented in figure~\ref{f:pw-mach}a for the different NPR's. It can be noted the steep increase of the 
wall pressure due to the lambda foot of the separation shock, then the flat behavior in the separation bubble and finally a mild increase to the back pressure value in the final part of the diffuser.
The isentropic Mach number distribution in the streamwise direction is reported in Figure~\ref{f:pw-mach}b. This value is computed from the  
isentropic relation between the chamber pressure and wall pressure $p_0/p_w=\left(1+\frac{\gamma-1}{2} M_i^2  \right)^{\frac{\gamma}{\gamma-1}}$ and it is used to characterize the shock intensity 
in nozzle separated flows and to reduce the data from different experiments~\cite{Stark2005}.
The isentropic Mach numbers characterizing the shock intensities for the test cases with $NPR$ = 1.39, 1.46, and 1.54 are $M_i$ = 1.34, 1.41, and 1.52.
The lower value is close to the experimental value of the work of Bogar et al\cite{Corp1981}, 
while the other two are higher. It must be noted that Bogar, in order to characterize the shock intensity, employed the local Mach number at the edge of the top wall boundary layer immediately upstream of the shock,  instead of the  isentropic Mach number. Nonetheless, the two numerical values do not differ significantly. 

\begin{figure}
\centering
 {\includegraphics[width=0.60\textwidth]{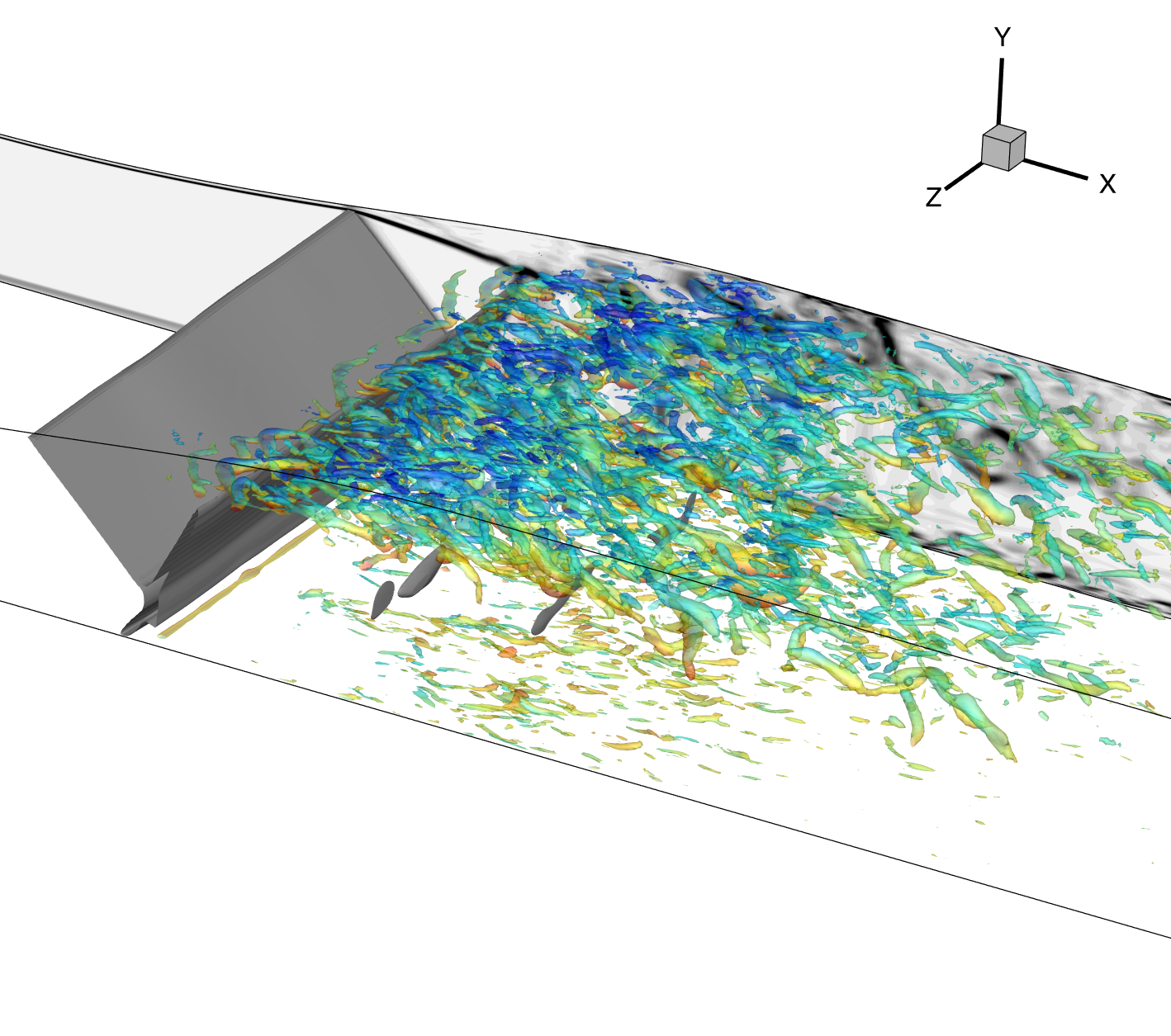}}
 \caption{ Iso-surface of the Q-criterion ($Q/(U/H_t)=60$), colored by the  local value of the streamwise velocity, for NPR = 1.54. The shock is visualized by an iso-surface of $\nabla \cdot u$; the 
           slice in the Z-plane shows the field of $||\nabla \rho||$.} 
 \label{f:iso-q}
\end{figure}

\begin{figure}
 \begin{subfigmatrix}{1}
  \subfigure[]{\includegraphics[width=0.70\textwidth]{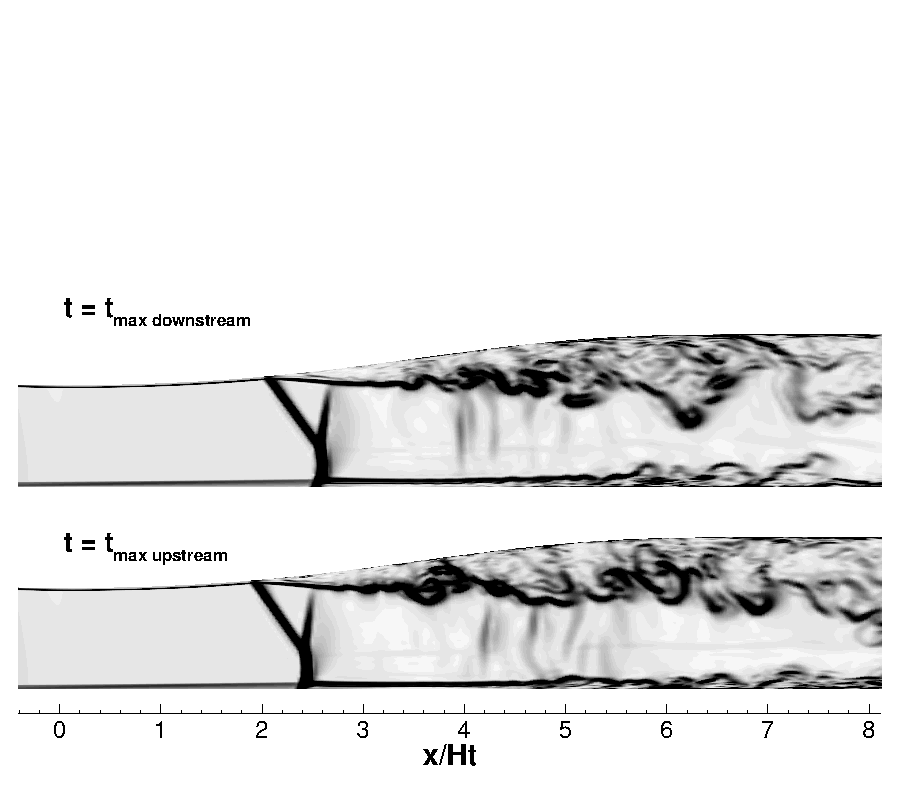}}
 \end{subfigmatrix}
 \caption{Numerical Schlieren at two different time instants for NPR = 1.39.}
 \label{f:up_down}
\end{figure}

The main characteristics of the instantaneous flowfield are shown in figure~\ref{f:iso-q} for NPR=1.54. The turbulent structures are represented by 
showing a positive iso-value of the $Q$-criterion \cite{Dubief2000}. This qualitative criterion defines as vortex tubes the regions 
where the second invariant of the velocity gradient tensor $Q$ is positive:
$$Q = \frac{1}{2} (\Omega_{ij}\Omega_{ij}-S_{ij}S_{ij})>0  $$
where $S_{ij}$ and $\Omega_{ij}$ are the symmetric and anti-symmetric components of $\nabla u$. A value of $Q/(U/H_t)=60$  has been chosen and 
the iso-surface are colored by the local value of the streamwise  velocity. Is is possible to notice at first the roll-up of almost two-dimensional vortical structures in the shear layer, 
which are bent toward the direction of motion and are rapidly replaced by three-dimensional structures developing downstream. 
In addition, a global unsteadiness with fluctuations in the separation shock position 
characterizes the flowfield, as shown in figure~\ref{f:up_down}, where two different snapshots of the density gradient field,  showing the extreme position reached by the shock system, are reported for NPR = 1.39.   
The figure also highlights the early development of shear layer instabilities downstream the separation line, thanks to the defition given by Eq.~\ref{eq:delta} for the subgrid length-scale $\Delta$.

%


\subsection{Wall pressure signature}
\label{sec:WallPSig}
\begin{figure}
 \begin{subfigmatrix}{2}
  \subfigure[]{\includegraphics[width=0.40\textwidth]{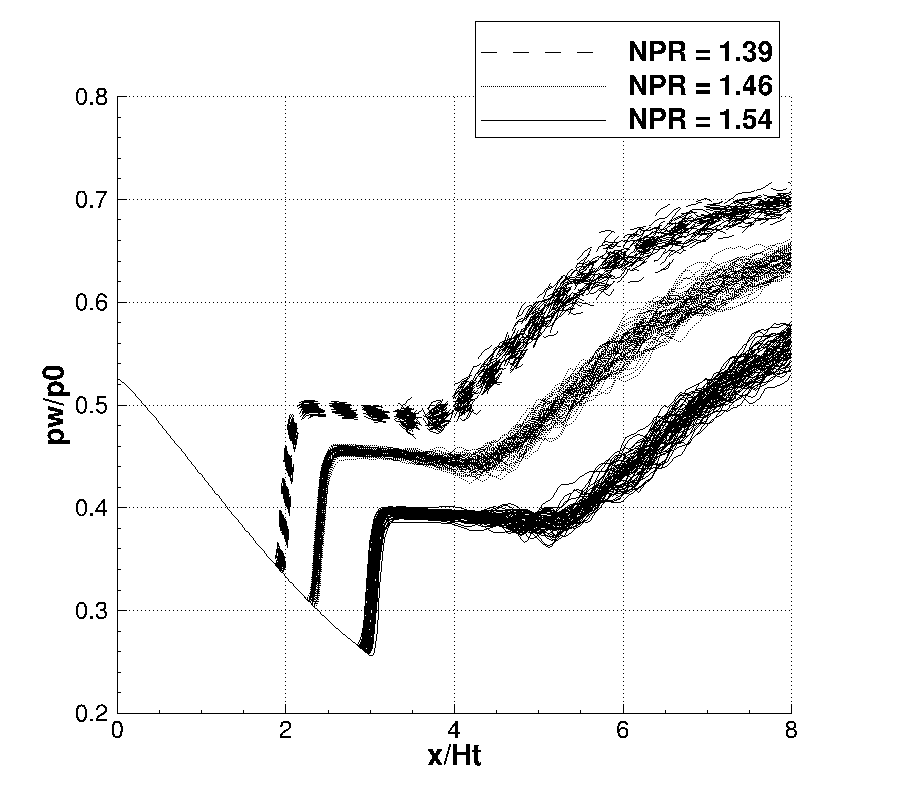}}
  \subfigure[]{\includegraphics[width=0.40\textwidth]{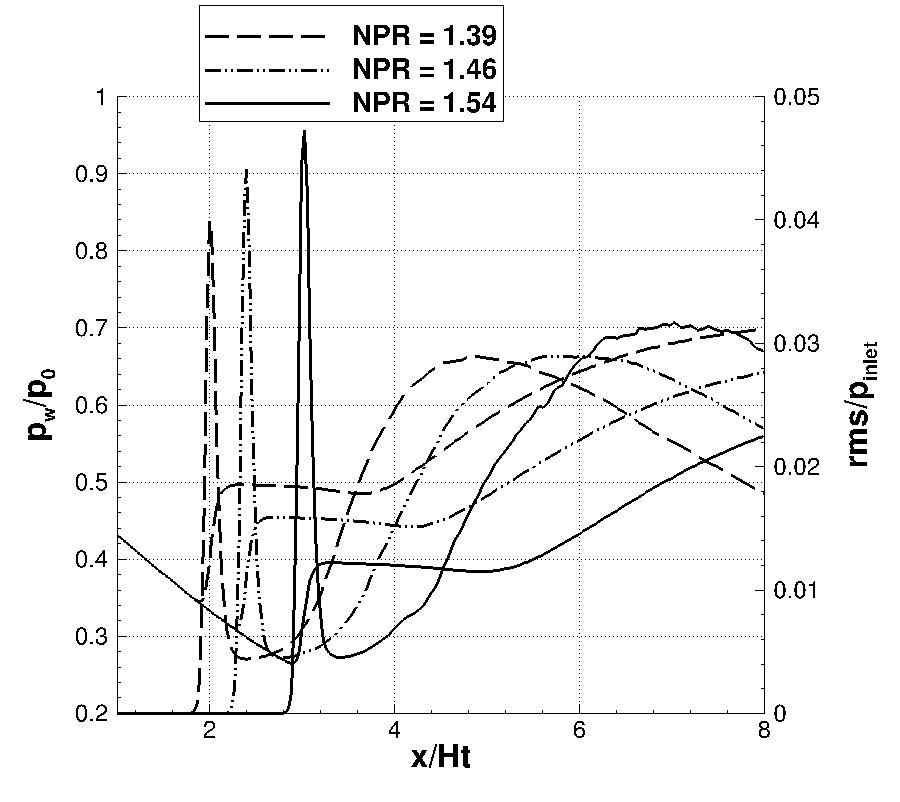}}
 \end{subfigmatrix}
 \caption{a): Streamwise distributions of instantaneous spanwise averaged wall pressure; b): streamwise distributions of time and spanwise averaged wall pressure and the pressure root mean square.} 
 \label{f:pw}
\end{figure}

\begin{figure}
 \begin{subfigmatrix}{2}
  \subfigure[]{\includegraphics[width=0.40\textwidth]{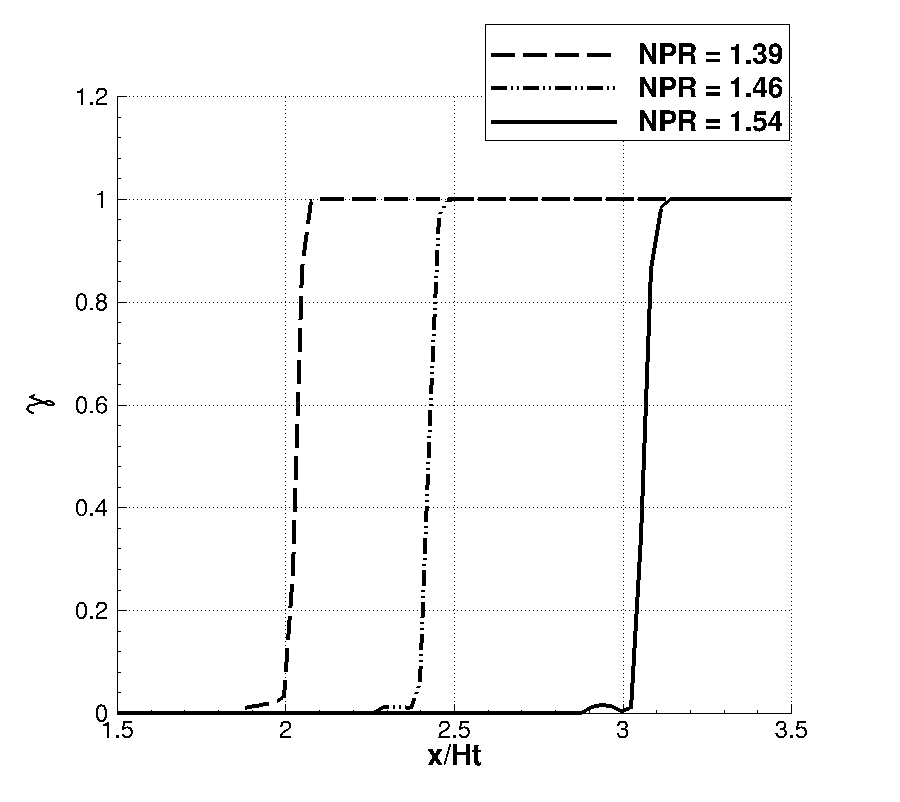}}
  \subfigure[]{\includegraphics[width=0.40\textwidth]{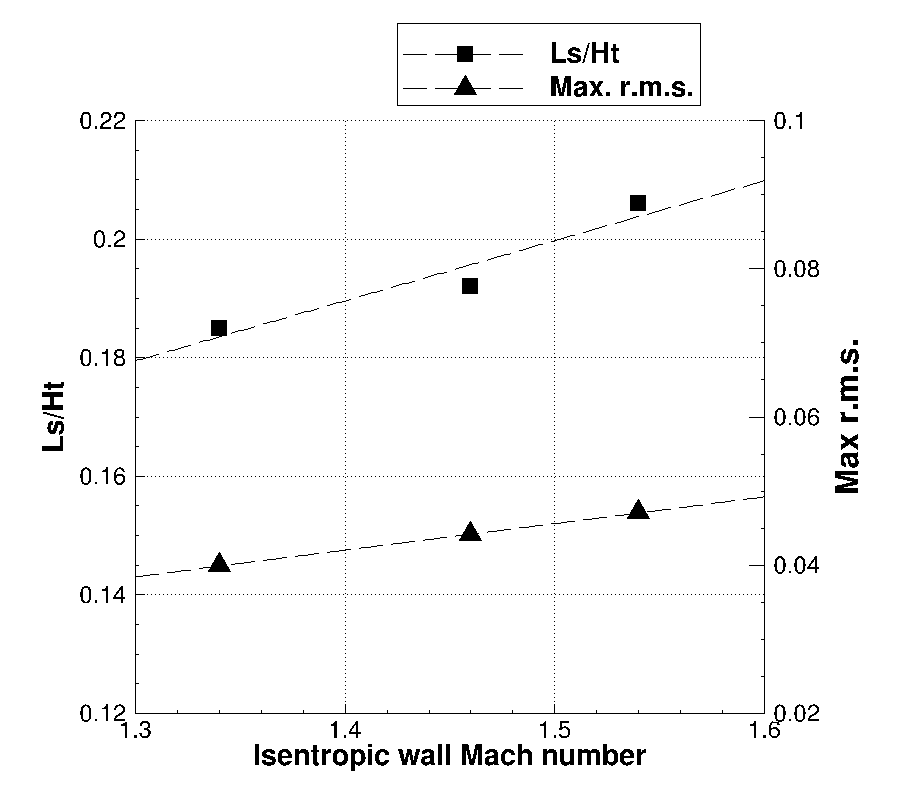}}
 \end{subfigmatrix}
 \caption{a): streamwise distribution of the intermittency factor. b): streamwise length of the shock motion $L_s$ and maximum value of the pressure r.m.s. as a 
function  of the wall isentropic Mach number.}
 \label{f:interm-mach}
\end{figure}

In this section the statistical properties of the fluctuating wall pressure are analyzed by evaluating the root mean square and the intermittency factor.
Figure~\ref{f:pw}a shows a set of instantaneous wall pressure distributions and illustrates the entity of the shock excursion. The root mean square value (r.m.s) 
of the top pressure fluctuations is reported in figure~\ref{f:pw}b. Within the attached boundary layer the r.m.s. of the top pressure fluctuations is zero, since, according to the DDES approach, this flowfield region is automatically treated in RANS mode. 
Instead, downstream of the separation point, there is a sharp peak in the r.m.s. value,  corresponding to the excursion zone of the shock system. Moving downstream, the 
first part of the recirculation region is characterized by a decrease of the r.m.s. of approximately one order of magnitude, while  a mild increase of the r.m.s. is observable in the last part of the diffuser.
The maximum value of the r.m.s. increases almost linearly with the isentropic Mach number, as shown in figure~\ref{f:interm-mach}b. 
It can be noted that the distribution of the r.m.s. of the wall pressure fluctuations is qualitatively very similar to the distributions found in other classical shock wave/boundary layer interaction; see for example the experimental findings of Dupont on an incident shock on a flat plate~\cite{Dupont2006} and 
of Dolling on a supersonic flow over a compression ramp~\cite{Dolling1985}.


In the region where the wall pressure $p_w$ is intermittent, an intermittency factor $\gamma$ can be defined~\cite{Dolling1985}, this representing the fraction of the time 
that $p_w$ is above the maximum pressure of the attached boundary layer, i.e.,
$$\gamma = \textrm{time}[p_w > (\bar{p_w}+3\sigma_w)]/\textrm{total time}$$
For the undisturbed boundary layer, the experimental value of $\gamma$ is equal to 0.0015 (close to the theoretical Gaussian value of 0.0013). In the RANS 
simulated attached boundary layer the value of $\gamma$ would be zero. 
An intermittency equal to 0.5 corresponds to case of having the same probability for the shock to be located on the left and on the right of the probe, and coincide with the maximum value in the r.m.s. distribution. 
The streamwise evolutions of the intermittency for the various NPR's are shown in figure \ref{f:interm-mach}a. All the distributions of the intermittency versus $x/H_t$ have 
the same shape, with the value of 0.5 occurring at the same abscissa of the maximum r.m.s. value. This shape is similar to the one produced by the shock in a supersonic ramp flow (Dolling et al. \cite{Dolling1985}). 
The distribution of $\gamma$ can be used to evaluate the shock excursion length. In fact, at any instant, the furthest upstream position of the separation shock is where the 
incoming boundary layer is firstly disturbed. Thus the distance over which $\gamma$ increases from 0.0015 (from zero in the present simulations) to 1 is 
the absolute length scale $L_s$ of the shock motion. Figure~\ref{f:interm-mach}b shows the nondimensional shock excursion length scale $L_s/H_t$ as a function of the 
isentropic Mach number $M_i$. It can be seen that the trend is almost linear, with an increase of 11\% in the value of $L_s/H_t$ when rising the isentropic Mach number from 1.34 to 1.52.

%

\begin{figure}
 \begin{subfigmatrix}{1}
  \subfigure[NPR = 1.54]{\includegraphics[width=0.16\textwidth,angle=270,clip]{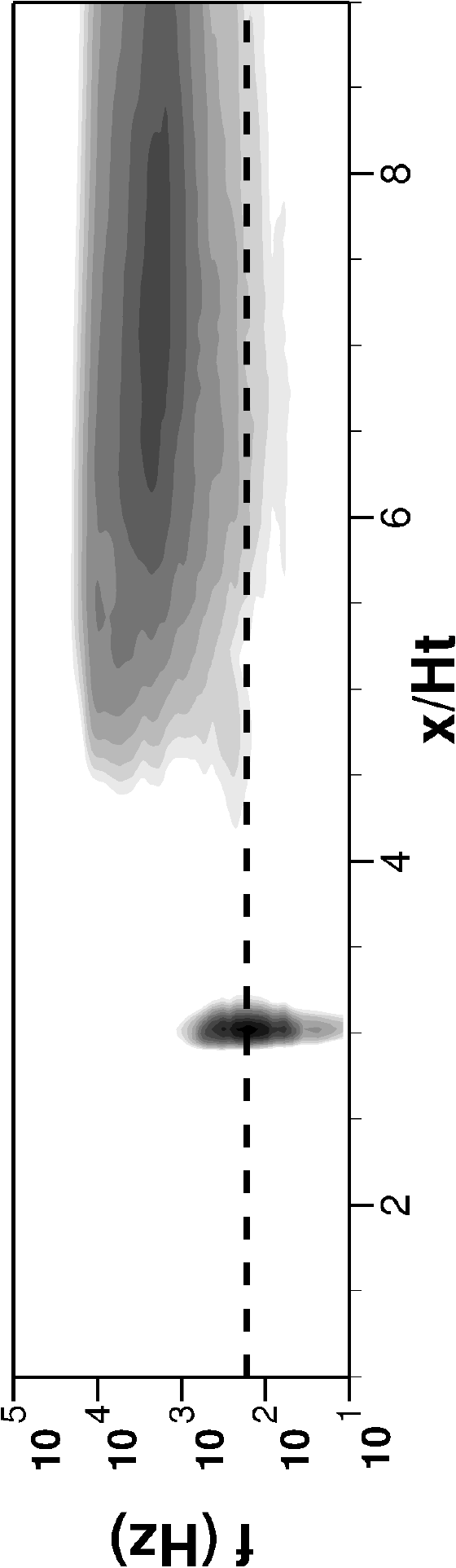}}
  \subfigure[NPR = 1.46]{\includegraphics[width=0.16\textwidth,angle=270,clip]{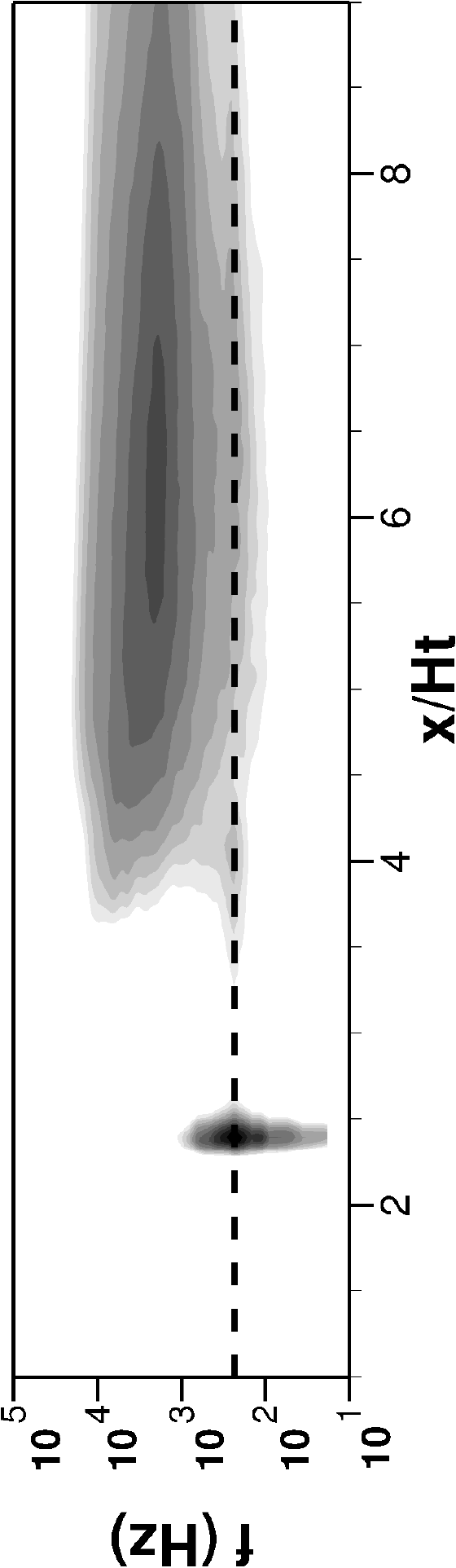}}
  \subfigure[NPR = 1.39]{\includegraphics[width=0.16\textwidth,angle=270,clip]{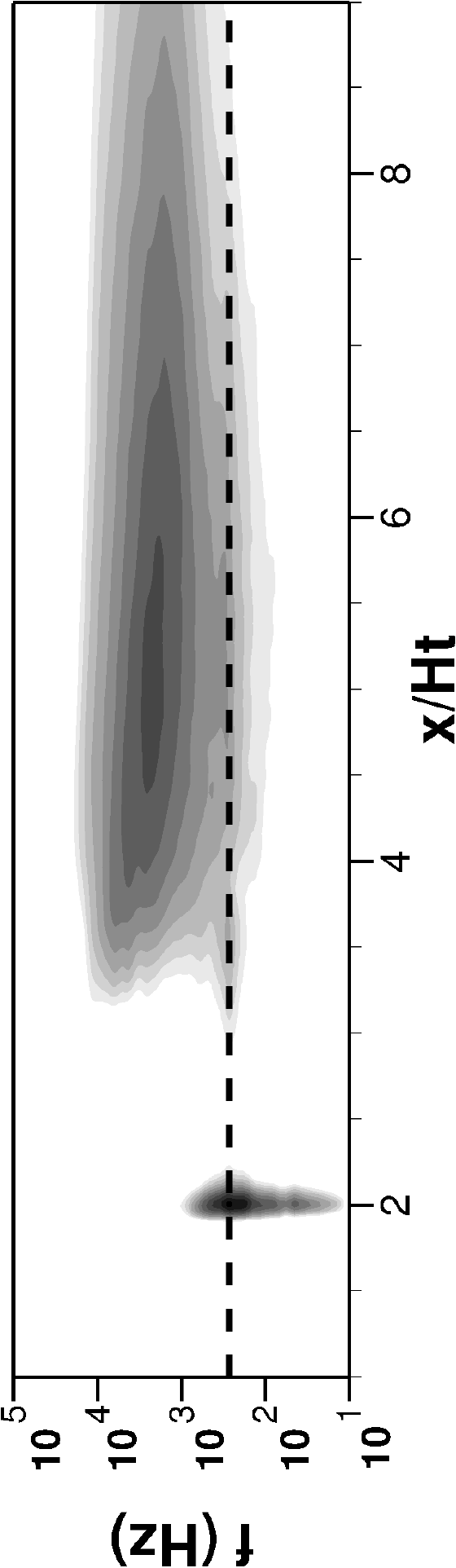}}
 \end{subfigmatrix}
 \caption{Pre-multiplied spectra ($f \, E(f)$) of the top wall pressure at various NPR's.
 The horizontal dashed line denotes the peak frequency of the shock motion.}
 \label{f:fft_npr}
\end{figure}

The pre-multiplied spectra $f \, E(f)$ of the pressure signal at the top wall
are shown in figure~\ref{f:fft_npr} for the different NPR's as a function of
the dimensional frequency $f$ and the streamwise coordinate $x/H_t$.
The power spectral densities have been computed using the Welch method,
subdividing the overall pressure record into $K$ segments with 50\% overlapping,
which are individually Fourier-transformed.
The frequency spectra are then obtained by averaging the periodograms of the various segments,
which allows to minimize the variance of the PSD estimator, and by applying a Konno-Omachi
smoothing filter~\cite{Konno1998}, which ensures a constant bandwidth on a logarithmic scale.
The number of segments is $K=10$ for the three cases here investigated.
The spectral maps are characterized by two different zones, qualitatively similar for the various NPR's.
The first region is associated with the dynamics of the shock system
and is identified by a peak whose characteristic frequency is of the order $O(200) Hz$,
located in the proximity of the shock foot. We point out that,
while previous investigations based on URANS identified the shock motion as tonal~\cite{Hsieh1987},
the low-frequency activity predicted by our DDES is rather broadband, the energy content 
encompassing a whole decade of frequencies. This behavior is in agreement with recent
experiments and LES carried out for canonical supersonic boundary layer interactions~\cite{Dupont2006,Aubard2013,Bermejo-Moreno2014,pirozzoli_10_1}.
The second extended region in the spectral densities is the signature of the turbulent activity in the separation bubble,
whose dynamics is well captured by the LES branch of the simulations. The peak of the frequency spectra in this zone is 
centered around $f \approx 2500 Hz$ and its streamwise location approximately correspond to the
reattachment point. 
This qualitative scenario is shared by the various NPR's. The main effect of increasing the NPR
is to shift downstream the location of the low-frequency peak and to (slightly) decrease the characteristic frequency of
the oscillations (see section~\ref{s:wavelet-results} for a comparison with experiments).

\subsection{Wavelet spectral analysis}
\label{sec:SpecAna}
\subsubsection{Morlet wavelet transform}
\begin{figure}
\centering
 {\includegraphics[width=0.60\textwidth]{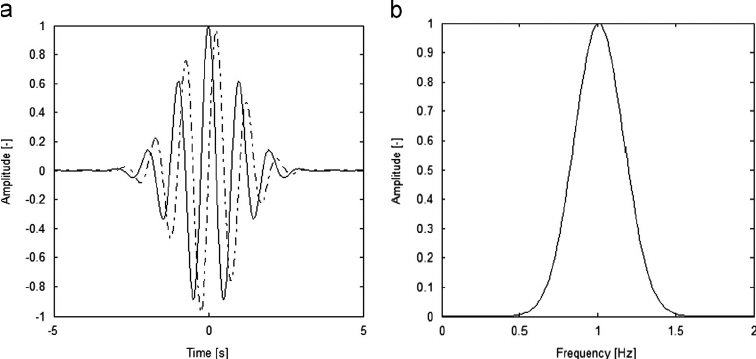}}
 \caption{ Morlet Wavelet base: a) real part (solid line) and imaginary part (dashed line) in the time domain; b) the corresponding wavelet in the frequency domain. } 
 \label{f:CM}
\end{figure}

The continuous wavelet transform is applied to the unsteady wall pressure signals in order to decompose them in the time-frequency space. 
An extended review of the application of wavelets to study turbulence phenomena can be found in Farge~\cite{Farge}, while only the key theoretical aspects are here reported.
The continuous wavelet transform of a discrete time sequence $p_n$, with equal spacing $\delta t$ and $n=0...N-1$, is defined as the convolution of $p_n$ with 
a scaled and translated version of the mother wavelet $\psi_0$:
\begin{equation}
W_n(s) = \sum_{n'=0}^{N-1} p_n \cdot \psi^{*}\left[\frac{(n'-n)\delta t}{s}\right] 
\label{eq:def}
\end{equation}
where $*$ denotes the complex conjugate. By varying the wavelet scale $s$ and translating along the time index $n$, one can construct a picture showing 
both the amplitude of any features versus the scale and how this amplitude varies with time. In this study, the Morlet wavelet has been chosen since higher 
resolution in frequency can be achieved when compared with other mother functions. It consists of a plane wave modulated by a Gaussian:
\begin{equation}
\psi_0 (\eta) = \pi^{-1/4} e^{i \omega_0 \eta} e^{-\eta^2/2} 
\label{eq:mother}
\end{equation}
where $\eta$ is a nondimensional time parameter and $\omega_0$ is the nondimensional frequency, here taken equal to 6 to satisfy the admissibility condition~\cite{Torrence1995}.
This wavelet is shown in figure~\ref{f:CM} both in the time and frequency domains.
From the definition of the wavelet coefficient one can directly define the wavelet power spectrum (WPS) as $|W_n(s)|^2$. The total energy is conserved under the wavelet transform and the equivalent of the 
Parseval's theorem for wavelet analysis is
\begin{equation}
\sigma^2 = \frac{\delta j \delta t}{C_{\delta}N} \sum_{n=0}^{N-1}\sum_{j=0}^{J}\frac{|W_n(s)|^2}{s_j}
\label{eq:power}
\end{equation}
where $\sigma^2$ is the variance, $\delta j$ is the scale spacing and $C_{\delta}$ is a factor coming from the reconstruction of a $\delta$ function from its wavelet transform. For 
more details the interested reader could see the work of Torrence and Compo~\cite{Torrence1995}. The energy density is then determined as:
\begin{equation}
E(s,t)= \frac{|W_n(s)|^2}{s_j}
\label{eq:density}
\end{equation}

Once a wavelet function has been chosen, it is necessary to determine a set of scales $s$ to use in the transform. In the case of non orthogonal wavelet analysis, it is possible 
to use an arbitrary set of scales to build up a more complete picture. Generally, it is convenient to write the scale as a fractional powers of two:
\begin{equation}
s_j=s_0 2^{j\, \delta j},~~ j=0,1,...,J
\label{eq:scale} 
\end{equation}
where $s_0$ is the smallest resolvable scale and J determines the largest scale. 
The scale $s_0$ should be chosen so that 
the equivalent wavelet period is approximately equal to $2\delta t$. The relationship between the equivalent Fourier period $\lambda$ and the wavelet scale $s$ can be found 
analytically~\cite{Torrence1995}. For the Morlet wavelet with $\omega_0=6$ it is possible to find that $\lambda = 1.03 s$, therefore they are almost equal. 
In the present analysis, the following parameter values have been chosen: $\delta t = 5\cdot 10^{-5} s$, $s_0=\delta t$, $\delta j=0.125$ and 
$J = 88$.

\subsubsection{Results of the wavelet analysis}
\label{s:wavelet-results}
\begin{figure}
 \begin{subfigmatrix}{2}
  \subfigure[]{\includegraphics[width=0.40\textwidth]{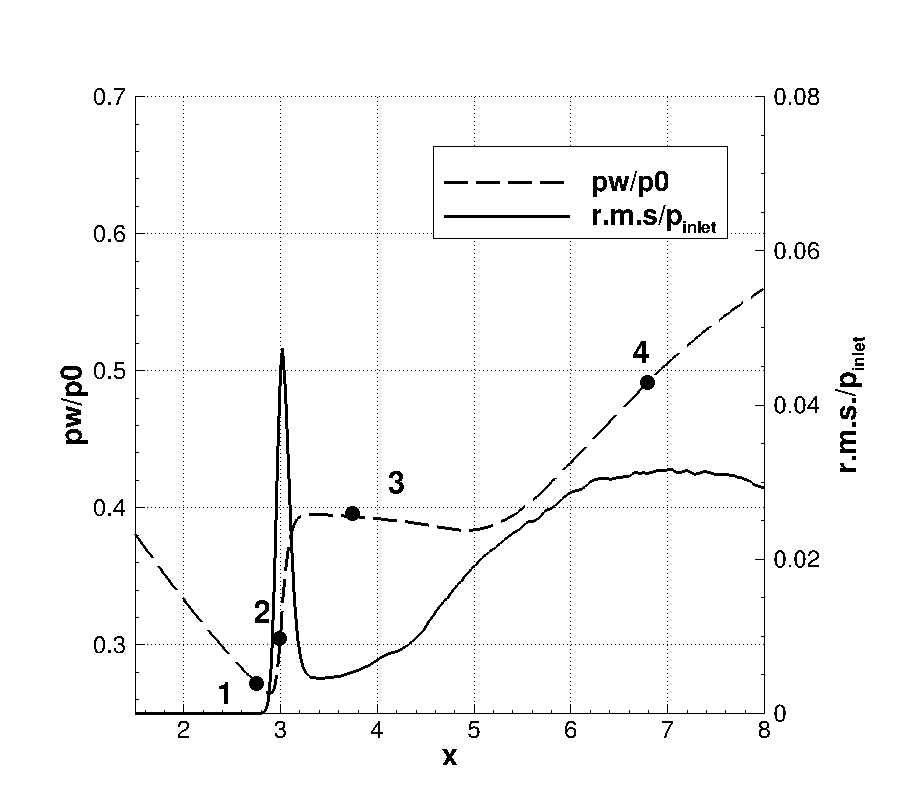}}
  \subfigure[]{\includegraphics[width=0.40\textwidth]{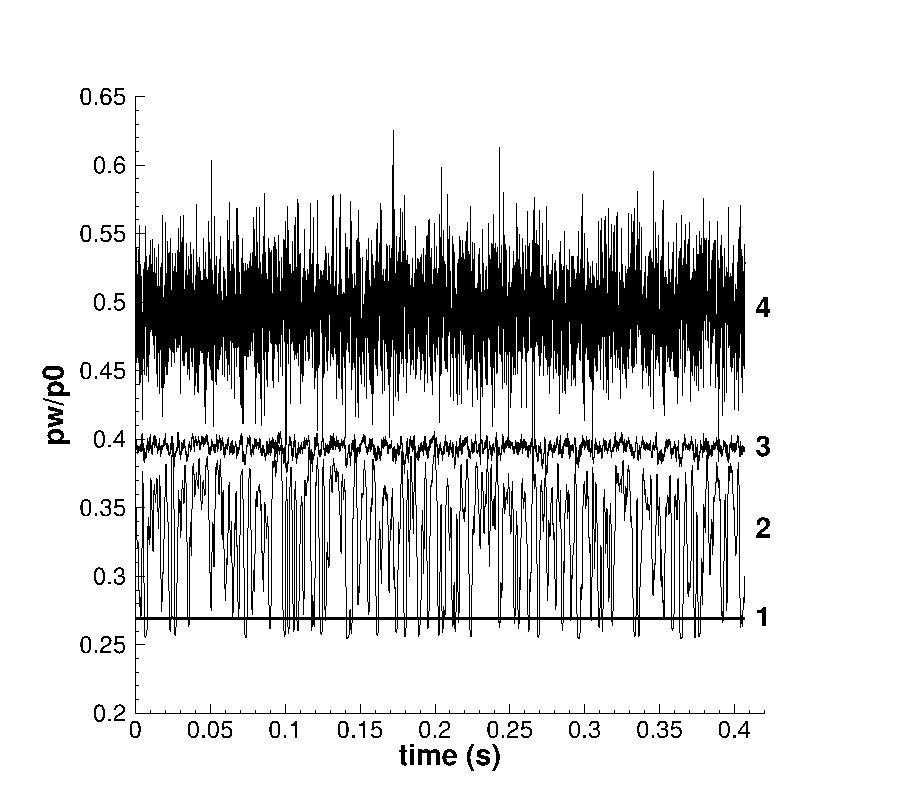}}
  \subfigure[]{\includegraphics[width=0.40\textwidth]{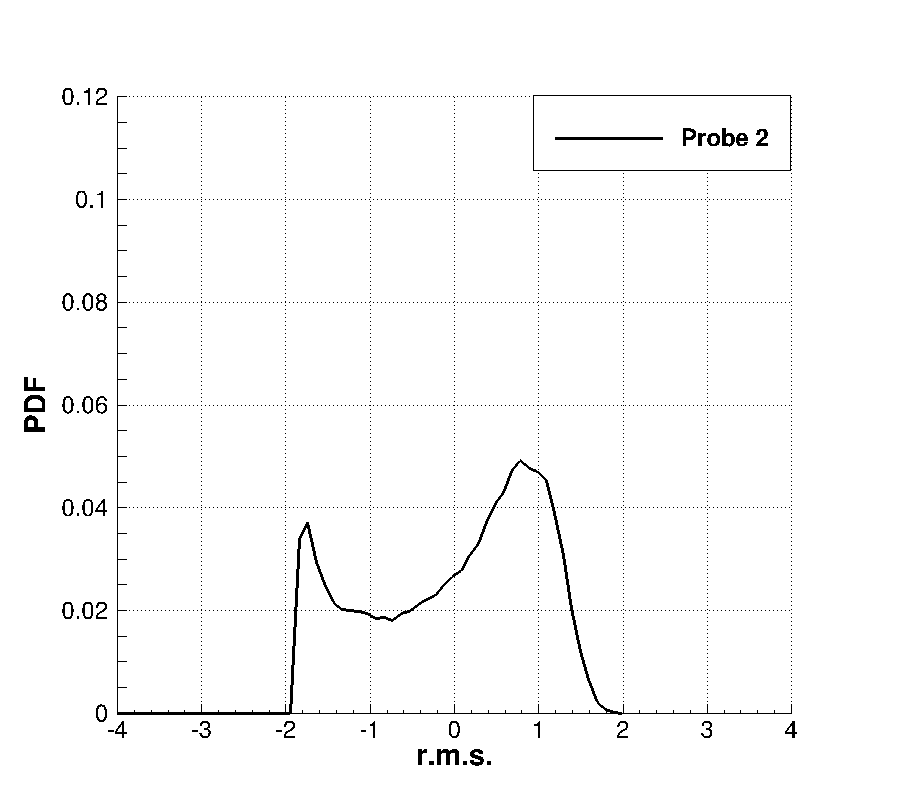}}
  \subfigure[]{\includegraphics[width=0.40\textwidth]{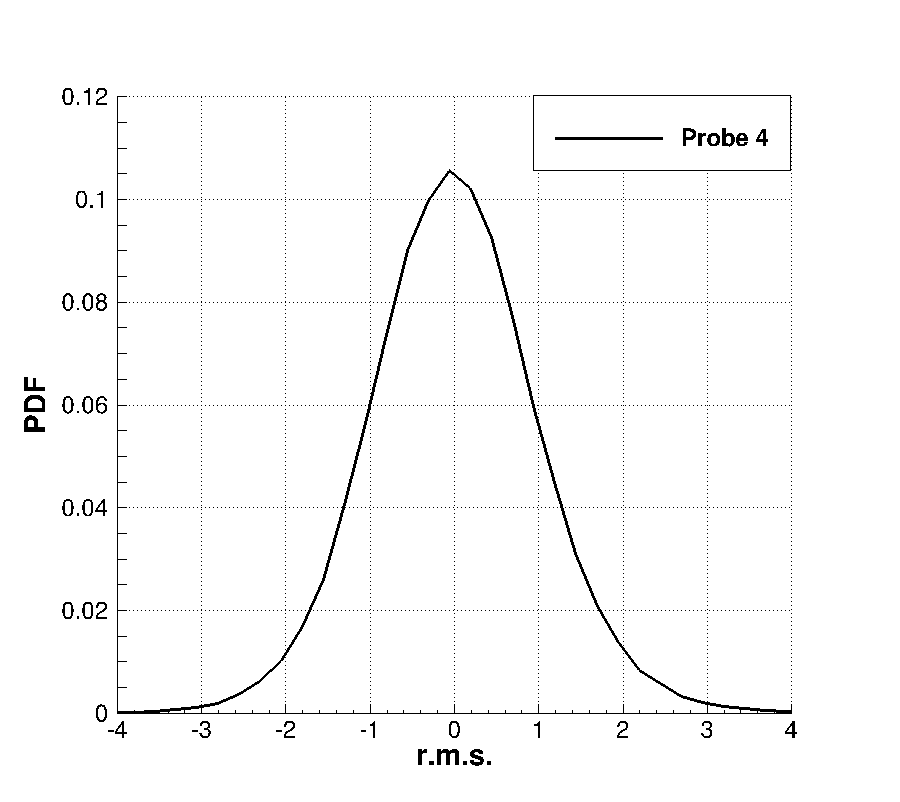}}
 \end{subfigmatrix}
 \caption{a): Streamwise distribution of time and spanwise averaged wall pressure with the root mean square for NPR = 1.53. The numbers indicate the pressure probes. b): intermittent pressure signals at probes from 1 to 4;
          c) probability density function at probe 2; d)  probability density function at probe  4. }
 \label{f:pwall-time}
\end{figure}

The time series of the fluctuating wall pressures are presented in figure~\ref{f:pwall-time}b for NPR = 1.54, being the results for the other NPR's very similar.
These signals are taken from the numerical probes displayed in figure~\ref{f:pwall-time}a. 
The first probe is located upstream the flow separation and its signal is almost constant in time, since this zone is the URANS domain (attached boundary layer).
The second probe is located in the region where there is the maximum value of root mean square of the pressure oscillation, that is the region of the shock excursion. 
As shown in figure~\ref{f:pwall-time}c,
the probability density function of wall pressures is bimodal, this being a characteristics of an intermittent signal. In facts, the wall pressure alternates between two different ranges: that of the attached boundary layer and that of 
the turbulent flow downstream of the separation shock, spending less time near the mean value which falls between the two extrema~\cite{Dolling1985}. Thus there are two maxima in 
the probability curve. The first peak is associated with the probability of finding $p_w$ in the narrow range of pressure associated with the attached boundary layer, 
hence showing a sharp peak. The latter peak has a broader maximum, that  reflects the probability of finding $p_w$ in the wider range of pressures that can be found downstream of the shock wave. 
The probe number 3 collects the signal at the beginning of the recirculating flow, which is characterized by a lower value of the oscillation amplitude with respect to the others. Finally, the
probe number 4 is located in the region of the vortex shedding, its signal shows a large oscillation amplitude and the probability density function is Gaussian, as shown in figure~\ref{f:pwall-time}.  

\begin{figure}
 \begin{subfigmatrix}{3}
  \subfigure[WPS/$\sigma^2$ at probe no. 2]{\includegraphics[width=0.30\textwidth]{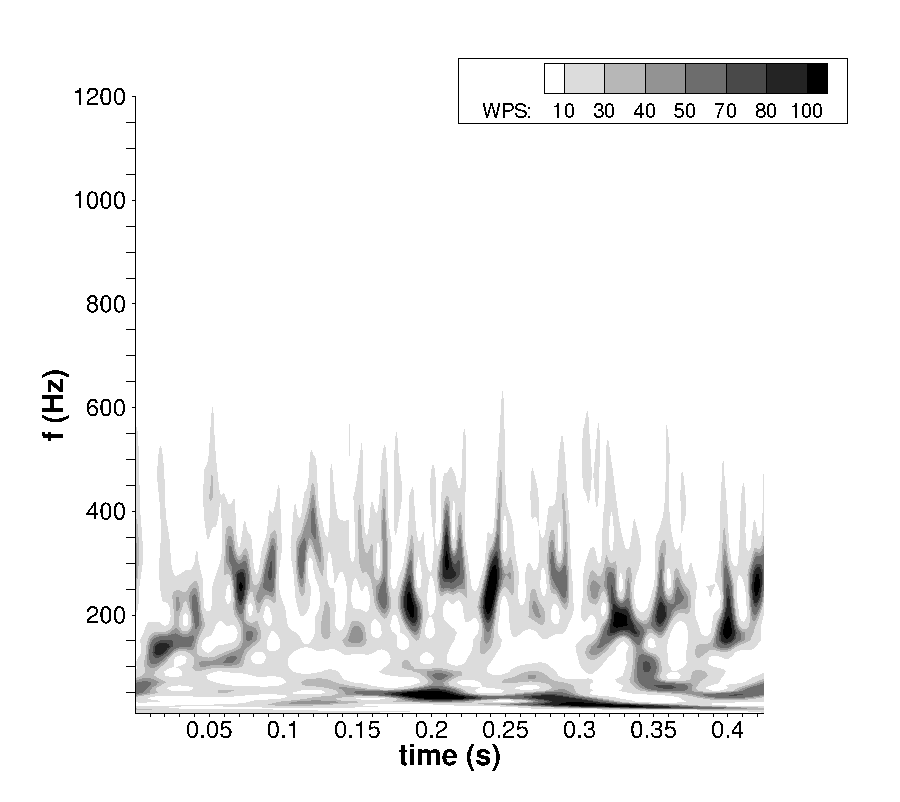}}
  \subfigure[Global WPS at probe no. 2]{\includegraphics[width=0.30\textwidth]{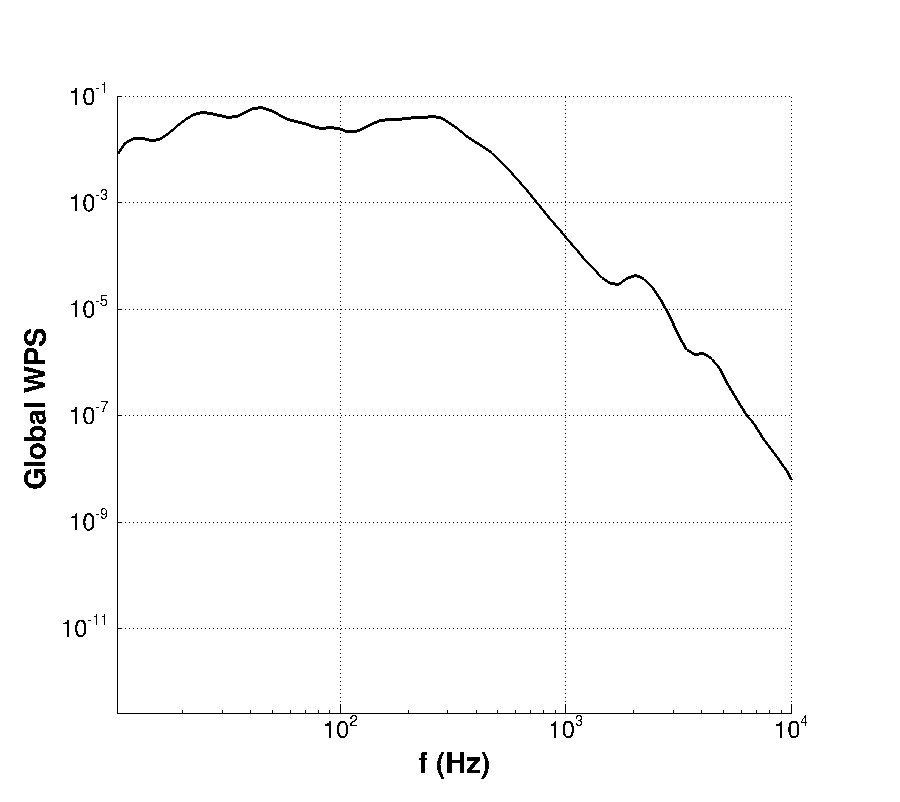}}
  \subfigure[Global WPS/scale at probe no. 2]{\includegraphics[width=0.30\textwidth]{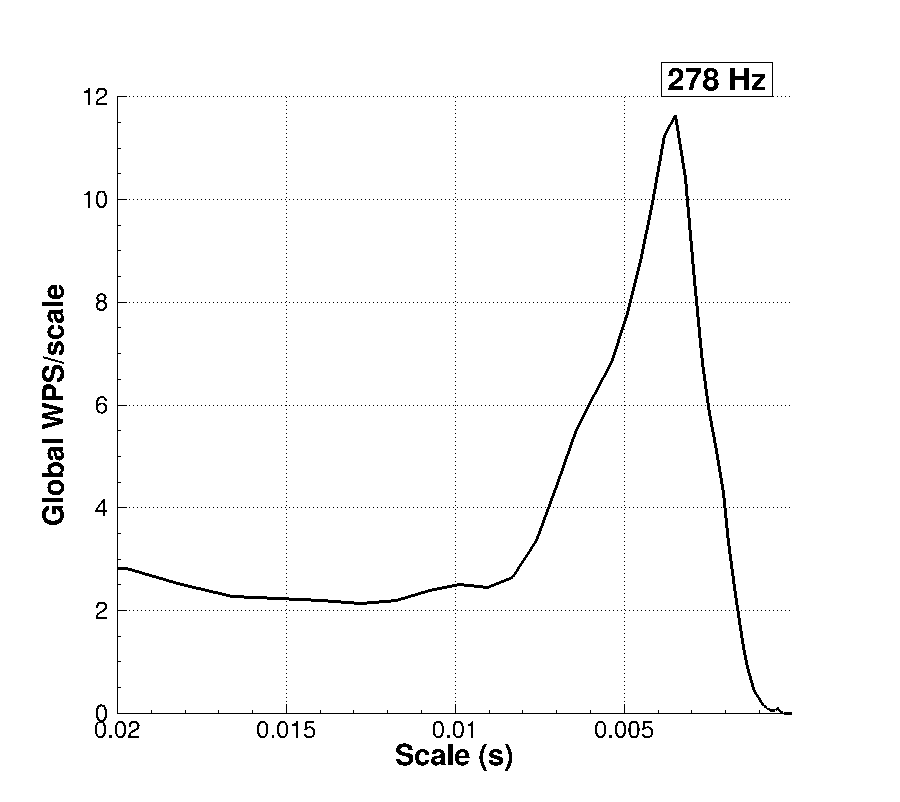}}
  \subfigure[WPS/$\sigma^2$ at probe no. 3]{\includegraphics[width=0.30\textwidth]{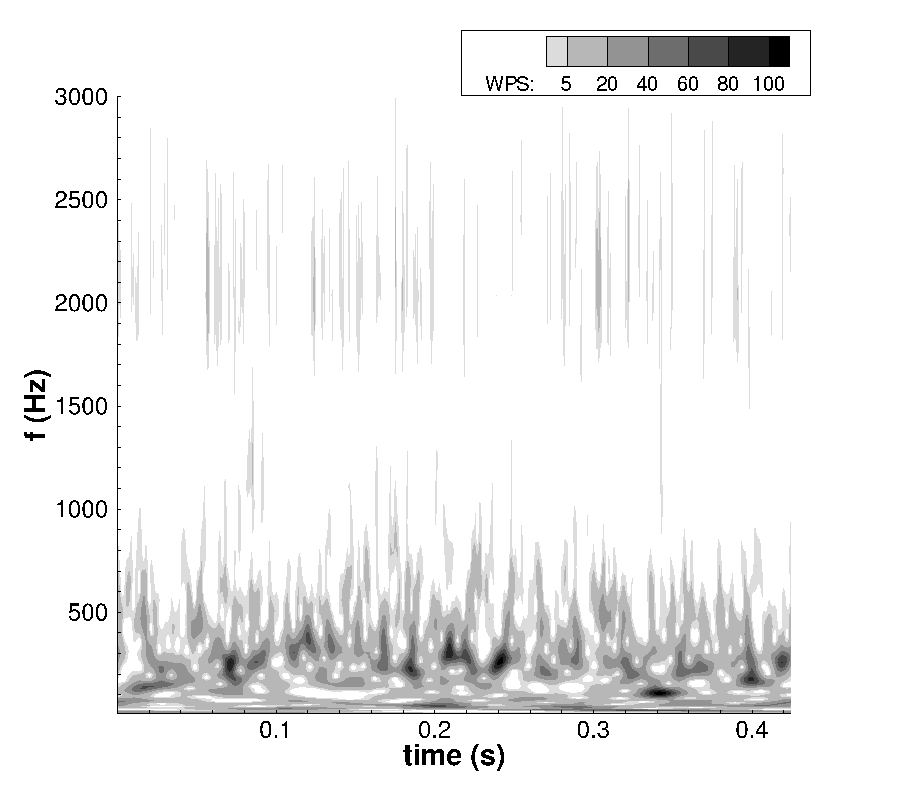}}
  \subfigure[Global WPS at probe no. 2]{\includegraphics[width=0.30\textwidth]{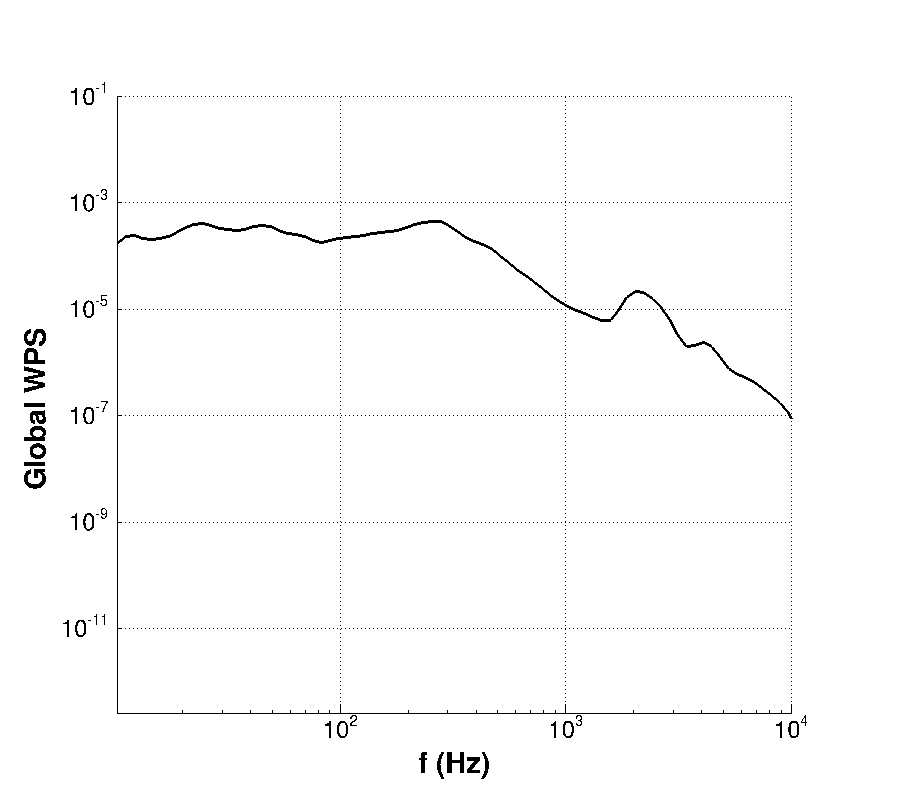}}
  \subfigure[Global WPS/scale at probe no. 3]{\includegraphics[width=0.30\textwidth]{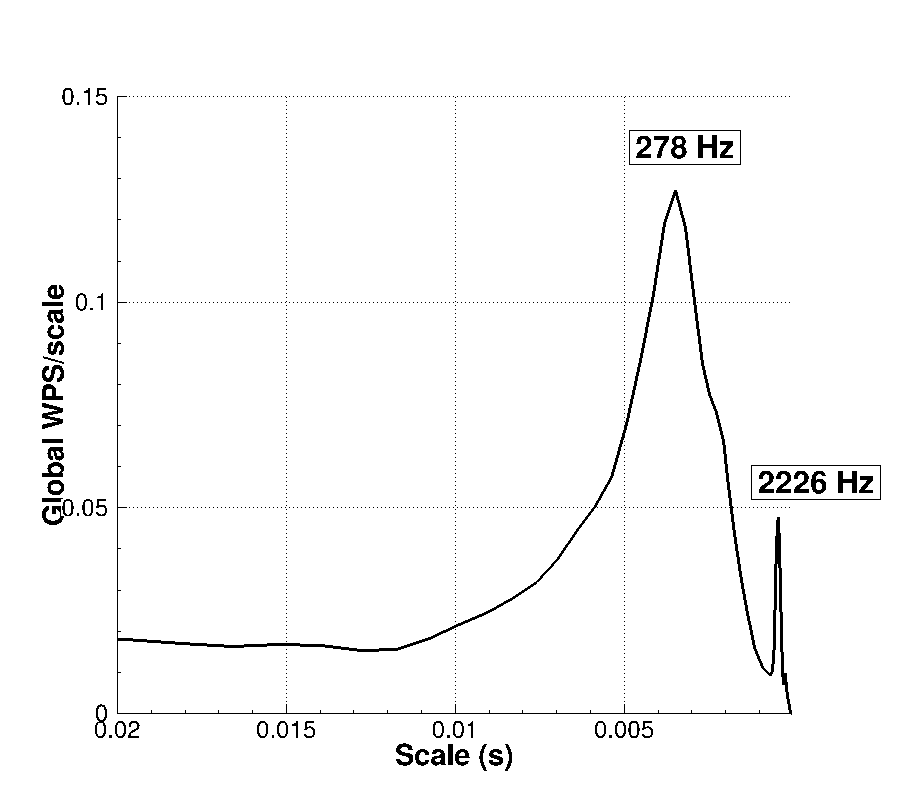}}
  \subfigure[WPS/$\sigma^2$ at probe no. 4]{\includegraphics[width=0.30\textwidth]{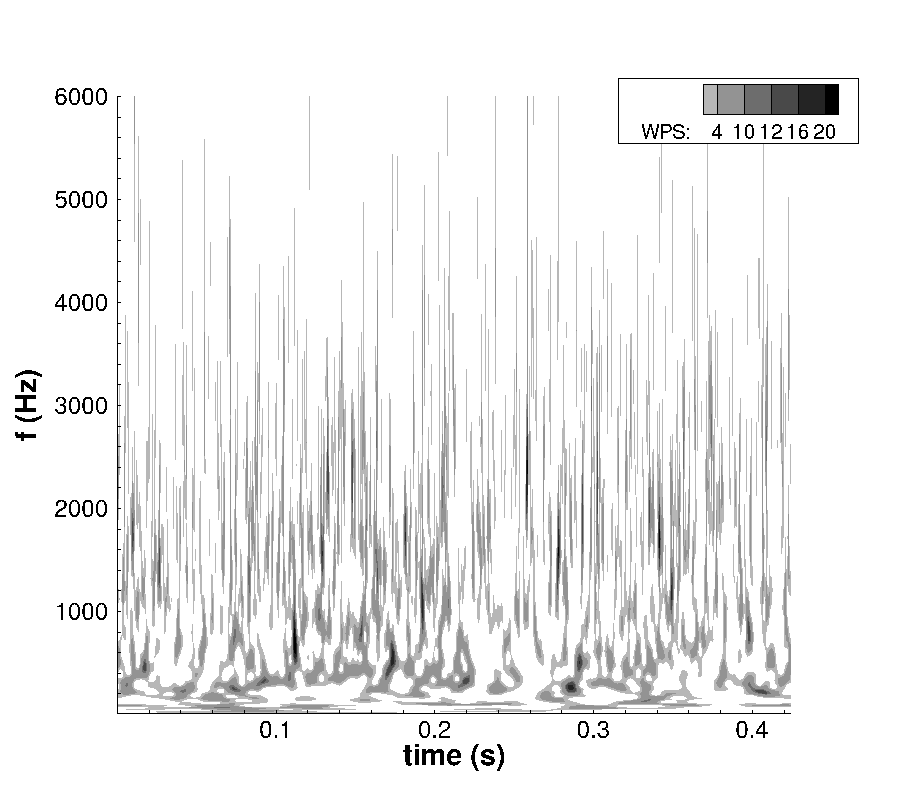}}
  \subfigure[Global WPS at probe no. 2]{\includegraphics[width=0.30\textwidth]{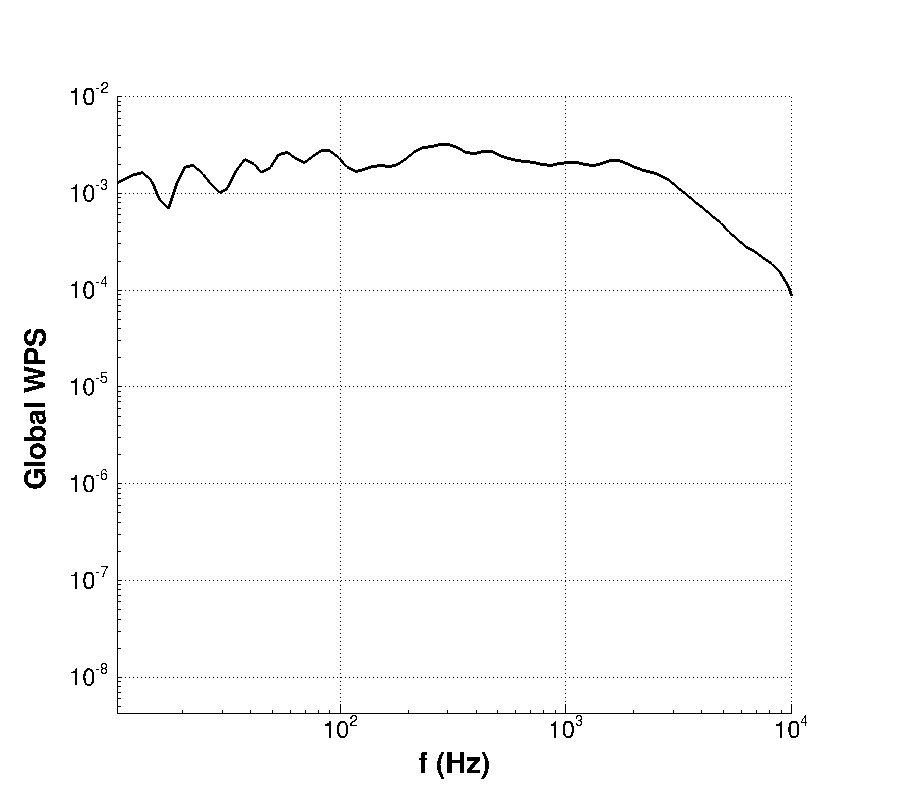}}
  \subfigure[Global WPS/scale at probe no. 4]{\includegraphics[width=0.30\textwidth]{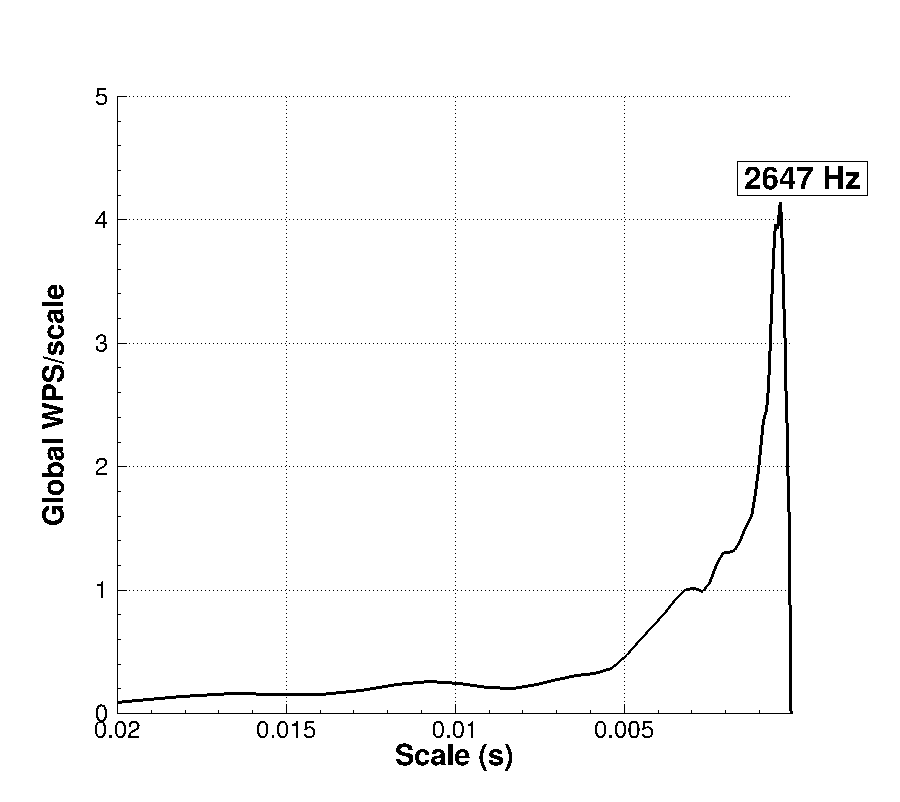}}
 \end{subfigmatrix}
 \caption{Wavelet Power Spectrum, global spectrum and global spectrum divided by the scale (energy density) of the pressure signals for NPR = 1.39.}
 \label{f:wps_npr_1}
\end{figure}

\begin{figure}
 \begin{subfigmatrix}{2}
  \subfigure[]{\includegraphics[width=0.40\textwidth]{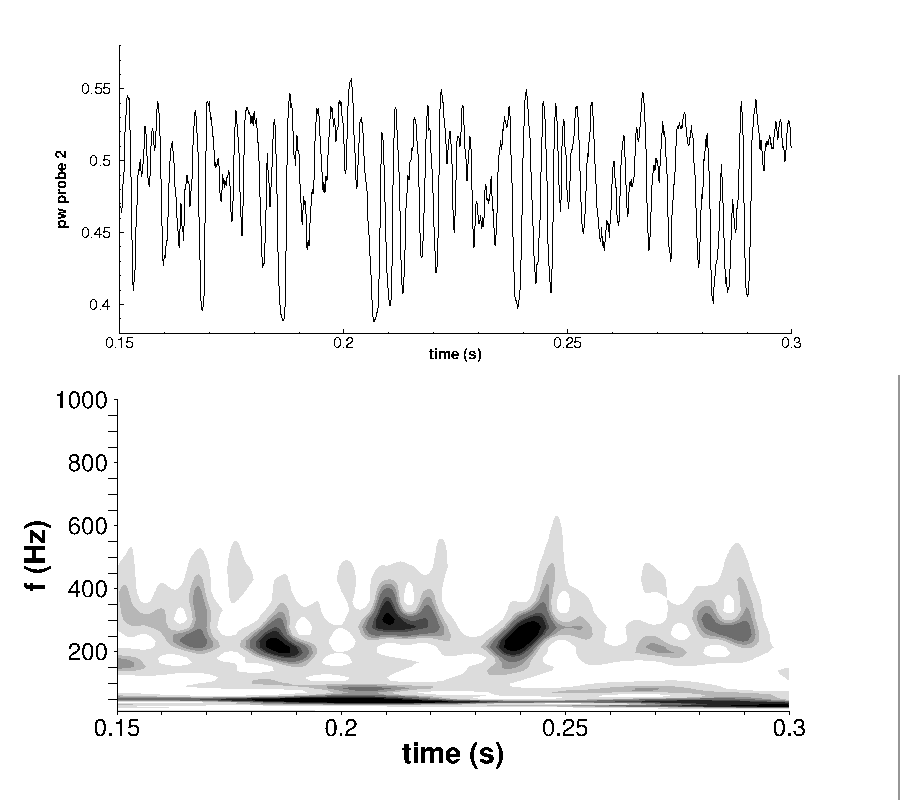}}
  \subfigure[]{\includegraphics[width=0.40\textwidth]{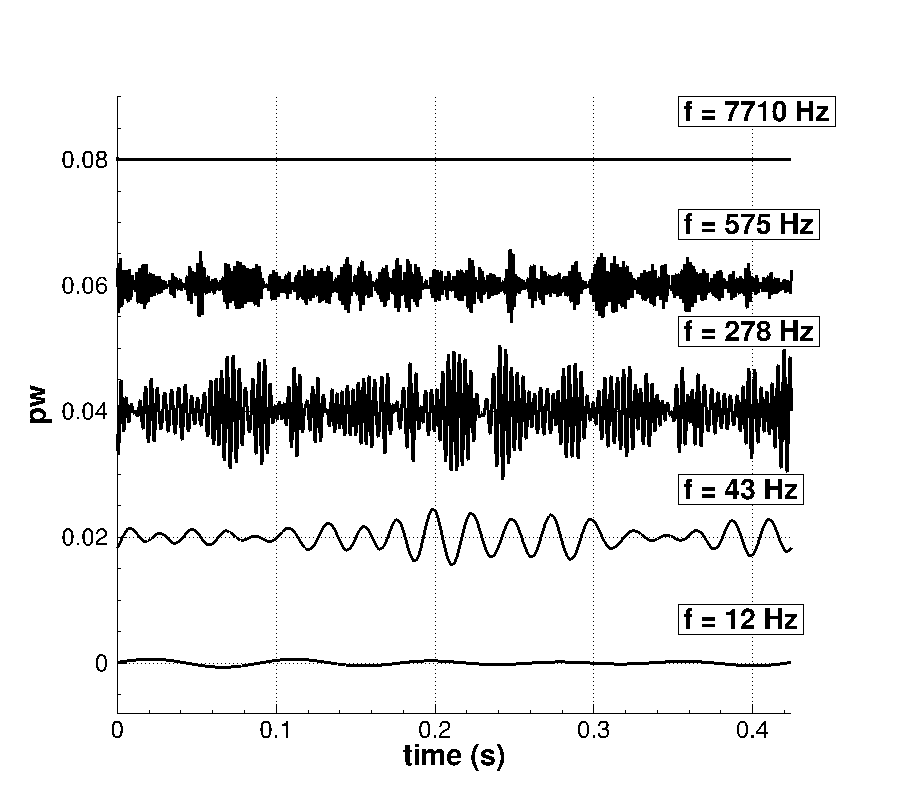}}
 \end{subfigmatrix}
 \caption{a): Wavelet Power Spectrum  together with wall pressure signal (probe no. 2) for NPR = 1.39.; 
          b): Time series for the wavelet coefficients of the pressure signal (probe no. 2) for different frequency modes. The amplitude of 
              the coefficients are shifted to facilitate the interpretation.} 
 \label{f:cfr_mode}
\end{figure}

The wavelet power spectrum of the wall pressure signals describes how the variance $\sigma^2$ of the wall pressure is distributed in frequency, as described by equation~(\ref{eq:power}).
Figure~\ref{f:wps_npr_1}a shows the normalized wavelet power spectrum $|W_n(s)|^2/\sigma^2$ in the frequency-time plane for the time series of the wall pressure from the second probe of the case at NPR = 1.39. 
The normalization by $1/\sigma^2$ gives a measure of the power relative to the white noise \cite{Torrence1995}.
The first aspect that can be extracted from this plot is that the spectrum appears as a collection in time of events, characterized by a variation of the  amplitude of the oscillation energy and a variation 
of the frequency of the most energetic events.  
For example, it is possible to see an important event at 0.18 s with a characteristic frequency around 220 Hz, 
then a second event at 0.21 s with an increase in frequency (around 300 Hz) and a third event at 0.24 s with a frequency of 250 Hz. 
Therefore it can be inferred from the data, that the shock movement is not continuous in time but rather intermittent. 
This aspect can be better appreciated in figure~\ref{f:cfr_mode}a which shows an 
enlargement of the wavelet power spectrum of the pressure signal from probe no. 2 between 0.15 s and 0.30 s, together with the pressure signal itself. It is evident from the picture that the 
pressure oscillation has an amplitude modulation, which is well captured by the wavelet power spectrum. 
To provide a better qualitative description of the dominant frequency modes within each of the main frequency branches, figure~\ref{f:cfr_mode}b shows the time series of a selection of the wavelet coefficients for different 
frequency modes. The time series, in fact, can be reconstructed by summing the real parts of the wavelet transform over all the scales:
\begin{equation}
p_w(n) = \frac{\delta j \delta t^{1/2}}{C_{\delta}\psi_0(0)}\sum_{j=0}^{J} \frac{\mathcal{R} [ W_n(s_j)]}{s_j^{1/2}}
\label{eq:recon} 
\end{equation}
From this picture it can be seen that the more relevant contributions come from frequencies between 50 Hz and 600 Hz, being the component at 278 Hz the most important. 
In addition, it is also possible to appreciate the amplitude modulation of the various components. 
These findings highlight the importance of an accurate time-frequency wavelet analysis in addition to the classical Fourier spectral analysis, since the energy and frequency fluctuations  
are not observable by means of the latter, that presents only time average information. 

\begin{figure}
 \begin{subfigmatrix}{3}
  \subfigure[]{\includegraphics[width=0.40\textwidth]{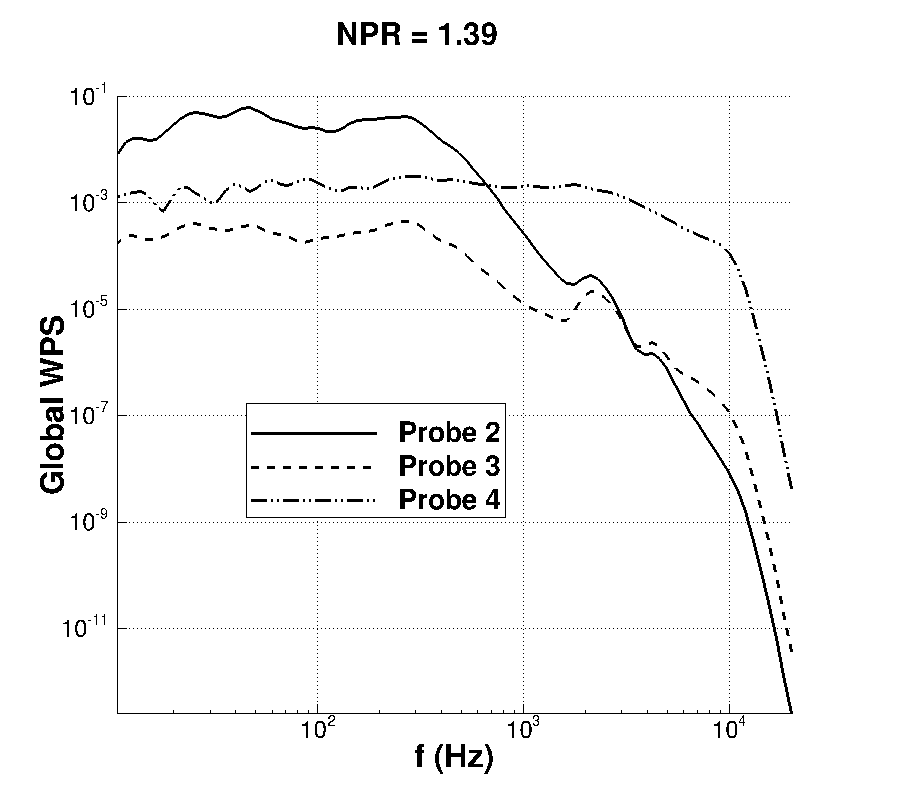}}
  \subfigure[]{\includegraphics[width=0.40\textwidth]{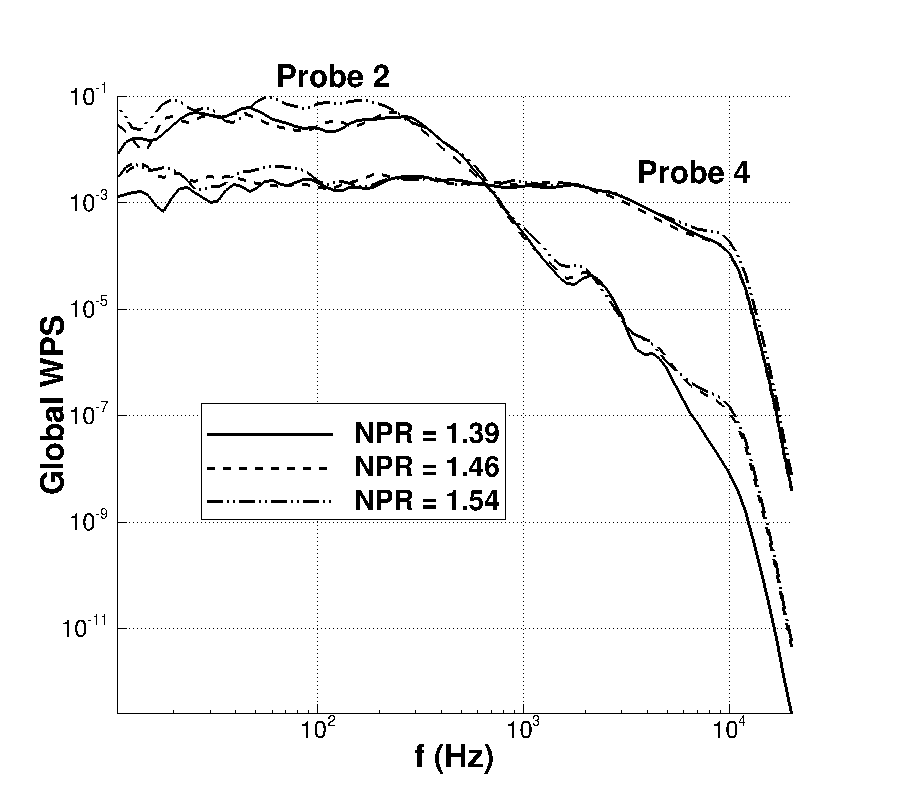}}
 \end{subfigmatrix}
 \caption{a): Comparison of the global WPS for the different probes at NPR=1.39; 
          b): Comparison of the global WPS for the different NPR's at probe no. 2 and 4.} 
 \label{f:cfr_wps}
\end{figure}

Figure~\ref{f:wps_npr_1}b shows the global wavelet power spectrum, that is the WPS integrated in time:
\begin{equation}
\overline{W^2}(s) = \frac{1}{N}\sum_{n=0}^{N-1} |W_n(s)|^2 
\label{eq:global} 
\end{equation}
while figure~\ref{f:wps_npr_1}c shows the global energy density $E(s)= \frac{\overline{W^2}(s)}{s_j}$ as a function of the scale. 
This last form is equivalent to the compensated spectra in the classical Fourier analysis. 
In this way,  it is possible to identify  the scales   most contributing to the energy, being possible to write~\cite{Torrence1995}:
\begin{equation}
\sigma^2 = \frac{\delta j \delta t}{C_{\delta}} \sum_{j=0}^{J}E(s) 
\label{eq:global-scale} 
\end{equation}
From figure~\ref{f:wps_npr_1}c it can be seen that there is a energy bump at large temporal scales (low frequencies), with 
a maximum  at 278 Hz. Therefore, the shock movement seems to be characterized by a broadband motion rather than by a sinusoidal motion. 
It may be worth full to recall that the frequency which gives the maximum value should be interpreted in a statistical sense, that is as the most probable frequency.
The analysis of the pressure signal from the probe 3 (located at the beginning of the recirculation bubble) is reported in figures~\ref{f:wps_npr_1}d,~\ref{f:wps_npr_1}e and ~\ref{f:wps_npr_1}f. 
Most of the energy is still located at low frequencies (lower than 1000 Hz) with an important bump at 278 Hz 
and a secondary bump at 2226 Hz. This second bump comes from integration in time of the intermittent events which can be seen in the frequency-time space 
between 2000 and 2500 Hz and it is linked to the vortex shedding of the shear layer. 
Figures ~\ref{f:wps_npr_1}g,~\ref{f:wps_npr_1}h and ~\ref{f:wps_npr_1}i represent the spectral analysis of the fourth probe, located in the vortex shedding region. 
The energy density now indicates that most of the energy is spread at higher frequencies, with a most probable frequency of 2647 Hz. In this region dominates the three dimensional vortical structures, even if there is still a energy contribution from the lower frequencies (below 1000 Hz).
Figure~\ref{f:cfr_wps}a compares the global wavelet power spectrum for all the pressure probes at NPR = 1.39, in order to quantify the shifting of the energy from the lower frequencies, characterizing 
the shock excursion region, to the higher frequencies which characterize the turbulent recirculating region. This picture is qualitatively the same shown by the Fourier analysis. 
The comparison of the global WPS for the different NPR's at pressure probes no. 2 and 4 are shown in figure~\ref{f:cfr_wps}b. It can be seen that, qualitatively, the 
behavior of the spectra are very similar. While, from a quantitative point of view, the signals from probe no. 2 show that NPR=1.54 has the highest power at low frequencies. 
This is correlated to the highest shock intensity of this NPR. The signals from probe no. 4, instead, are similar also from a quantitative point of view, indicating the same behavior for the 
turbulent separated region. 

\begin{figure}
\centering
 {\includegraphics[width=0.50\textwidth]{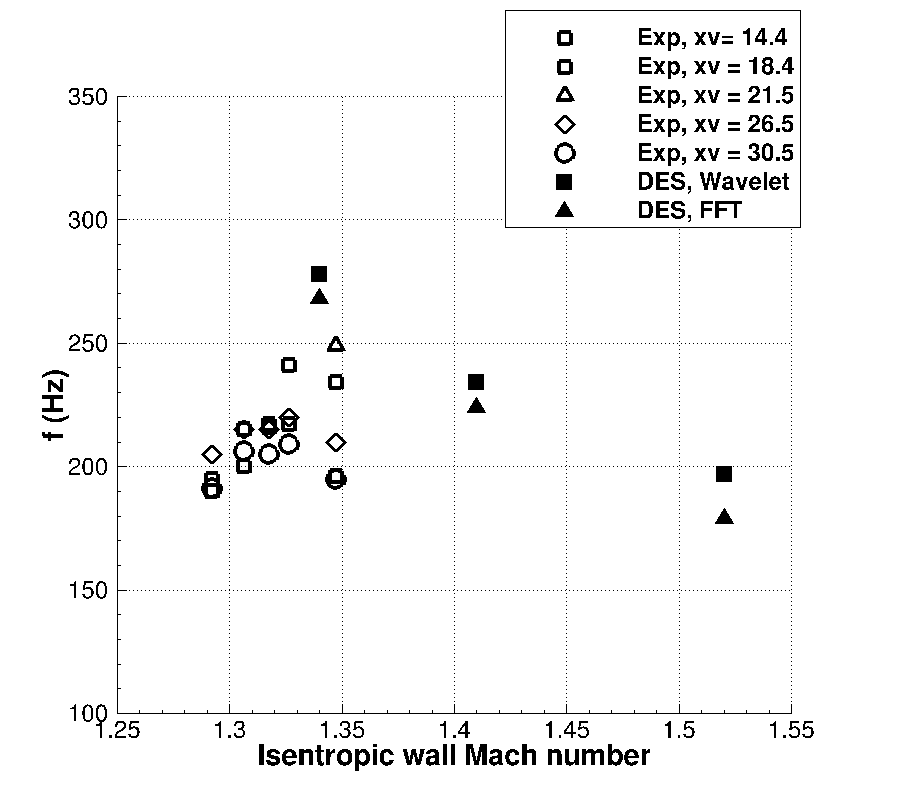}}
 \caption{Peak frequencies related to the shock movement as a function of the isentropic wall Mach number, computed with the wavelet analysis and the Fourier analysis.
 In the experimental data $xv$ is the non dimensional length of the nozzle.} 
 \label{f:peak-exp}
\end{figure}

Finally, figure~\ref{f:peak-exp} compares the values of the frequencies characterizing the shock movement  obtained with  wavelet and Fourier analysis with those taken from the Fourier analysis of the experimental data\cite{Corp1981} (open symbols). 
The test case with the isentropic Mach number equal to 1.34 is the only one which falls in the experimental range of shock Mach numbers~\cite{Corp1981} $M_s$ ($1.280 \le M_S \le 1.347$). 
The computed value result to be in reasonable agreement with the experimental ones, although the characteristic 
frequencies are sightly overestimated. 
We  can speculate that such discrepancy might be ascribed to some differences between the experiment and the simulations, as the presence of side walls and suction slots employed in the experimental configuration to remove the boundary layer.

\section{Conclusions}
\label{sec:Conclusions}
Delayed detached eddy simulations (DDES) of a planar nozzle with flow separation have been carried out for a Reynolds number, based on stagnation chamber properties and throat height, equal to $1.5 \cdot 10^6$ 
and for different nozzle pressure ratios (or equivalently different isentropic Mach numbers).
The nozzle flow simulated in this study is characterized by a separation shock with a classical lambda shape and by an important recirculation zone, which extends for several nozzle throat heights. All the simulations were 
able to capture a self-sustained unsteadiness of the shock system. 
As a first step, a classical statistical description of this unsteadiness has been carried out. The shock region is characterized by a well defined peak in the root mean square distribution of the 
oscillating wall pressure. The amplitude of this peak increases with increasing Mach number. The evaluation of the intermittency factor has allowed to evaluate the shock excursion length, which can reach the 
20\% of the throat height. All these findings qualitatively agree with the data of the experimental reference nozzle and with the data of other shock configurations, like  compression ramps 
and  incident shock waves on a flat plate. 
The spectral analysis has been conducted by using Fourier analysis and the Morlet wavelet transform, which is a well suited tool to analyze non stationary time series. 
The Fourier analysis has allowed to individuate a low frequency region, around 250 Hz, associated with the shock movement, and a higher frequency region (around 2500 Hz) associated with the turbulent separated flow. According to the wavelet analysis, 
the shock movement and the recirculating region  have been recognized to be characterized, in the time-frequency space, by a collection of events  with a modulation of the oscillation amplitude and a modulation of the frequency.

\section*{Acknowledgments}

The simulations have been performed thanks to computational resources provided by the Italian Computing center CINECA under the ISCRA initiative (grant IscrB\_SW-DES-1).
MB was supported by the SIR program 2014 (jACOBI project, grant RBSI14TKWU), funded by MIUR (Ministero dell'Istruzione dell'Universit\`a e della Ricerca).  
Wavelet software was provided by C. Torrence and G. Compo, and is available at URL: http://atoc.colorado.edu/research/wavelets/.





\bibliographystyle{elsarticle-num}
\bibliography{biblio}







%
%
%
%

\end{document}